\documentclass[eqn,opre, nonblindrev]{informs4}

\OneAndAHalfSpacedXI 

\usepackage{siunitx}
\usepackage{array}
\usepackage{booktabs}

\usepackage{graphicx}
\usepackage{tikz}
\usetikzlibrary{positioning, arrows.meta}
\usepackage{multirow}
\usepackage{hhline}
\usepackage{makecell}
\usepackage[small, margin=1cm]{caption}
\usepackage{comment}

\usepackage{appendix}

\usepackage{xcolor}
\definecolor{strcolor}{rgb}{0.6, 0.2, 0.6}
\definecolor{commentcolor}{rgb}{0.3125, 0.5, 0.3125}
\definecolor{keycol}{rgb}{0, 0, 1}

\usepackage{algorithm}
\usepackage{algpseudocode} 
\newcommand{\IndState}[1][1]{\State\hspace{\algorithmicindent}}

\newcommand{\oE}{\operatorname{E}}

\newcommand{\oP}{\operatorname{P}}
\newcommand{\oPr}{\operatorname{Pr}}

\newcommand{\mR}{\mathbb{R}}

\newcommand{\mN}{\mathbb{N}}

\usepackage{dutchcal}
\newcommand{\cF}{\mathcal{F}}
\newcommand{\cG}{\mathcal{G}}
\newcommand{\cO}{\mathcal{O}}

\newcommand{\mum}{\tilde{\mu}_{m}}
\newcommand{\pim}{\tilde{\pi}_{m}}

\newcommand{\tmn}{\tilde{\theta}_{m,n}}
\newcommand{\qm}{\tilde{q}_{m}}
\newcommand{\tm}{\tilde{\theta}_{m}}

\usepackage{bbm}
\usepackage{bm}

\usepackage{listings}
\usepackage{url}

\def\blot{\quad \mbox{$\vcenter{ \vbox{ \hrule height.4pt
				\hbox{\vrule width.4pt height.9ex \kern.9ex \vrule width.4pt}
				\hrule height.4pt}}$}}

\usepackage{natbib}
\renewcommand{\bibsection}{\section*{\refname}\vspace{10pt}}
\bibpunct[, ]{(}{)}{,}{a}{}{,}%
\def\bibfont{\fontsize{10}{10}\selectfont}%
\usepackage[colorlinks,citecolor=blue,urlcolor=blue,filecolor=blue]{hyperref}

\TheoremsNumberedThrough    
\ECRepeatTheorems

\EquationsNumberedThrough    
\gdef\AQ#1{}
\gdef\CQ#1{}

\newtheorem{alemma}{Lemma}[section]
\newtheorem{innercustomthm}{Lemma}
\newenvironment{customthm}[1]
  {\renewcommand\theinnercustomthm{#1}\innercustomthm}
  {\endinnercustomthm}

\begin{document}

\RUNAUTHOR{Lin, Song and Hong} %

\RUNTITLE{Efficient Nested Estimation of CoVaR}

\TITLE{Efficient Nested Estimation of CoVaR:\\ A Decoupled Approach}

\ARTICLEAUTHORS{

\AUTHOR{Nifei Lin}
\AFF{School of Management, Fudan University, Shanghai 200433, China}

\AUTHOR{Yingda Song}
\AFF{Antai College of Economics and Management, Shanghai Jiao Tong University, Shanghai 200030, China}

\AUTHOR{L. Jeff Hong}
\AFF{Department of Industrial and Systems Engineering, University of Minnesota, Minneapolis, MN 55455}

}

\ABSTRACT{
This paper addresses the estimation of the systemic risk measure known as CoVaR, which quantifies the risk of a financial portfolio conditional on another portfolio being at risk. We identify two principal challenges: conditioning on a zero-probability event and the repricing of portfolios. To tackle these issues, we propose a decoupled approach utilizing smoothing techniques and develop a model-independent theoretical framework grounded in a functional perspective. We demonstrate that the rate of convergence of the decoupled estimator can achieve approximately $\cO_{\rm P}(\Gamma^{-1/2})$, where $\Gamma$ represents the computational budget. Additionally, we establish the smoothness of the portfolio loss functions, highlighting its crucial role in enhancing sample efficiency. Our numerical results confirm the effectiveness of the decoupled estimators and provide practical insights for the selection of appropriate smoothing techniques.
}






\maketitle
\section{Introduction}

Financial institutions and markets are interconnected through various channels, including lending, trading, and derivatives. Fluctuations in risk factors such as stock prices, commodity prices, and interest rates can propagate through these channels, potentially leading to systemic risk. This occurs when the failure of a significant entity triggers a chain reaction, negatively impacting the entire financial system and the broader economy. The 2007–2009 financial crisis, for example, saw risks originating from structured investment vehicles spread across solvent institutions, ultimately sparking a global crisis. Given the potential consequences, understanding and managing systemic risk is crucial for maintaining financial stability and safeguarding the economy.

The first step of risk management is to timely monitor the risk through appropriate risk measures.
In this paper, we focus on the systemic risk measure CoVaR proposed by \cite{adrian2016covar}, which has been demonstrated to effectively capture the cross-sectional tail-dependence between financial institutions and has shown predictive capabilities regarding the 2007–2009 financial crisis. CoVaR is defined as the $\beta$-value-at-risk (VaR, also known as quantile) of a portfolio loss conditional on another portfolio loss being at its $\alpha$-VaR, where $\alpha,\beta\in(0,1)$ often take values such as $0.95$ or $0.99$ to represent the risk levels. 

Let $Z$ denote the realization of risk factors at a future time $\tau$, where $\tau$ is often one or two weeks from today and is the time horizon that the risk is assessed. Furthermore, let $\mu(z)$ and $\pi(z)$ denote the losses of the two portfolios respectively at time $\tau$ when $Z=z$. Notice that they are dependent on the realization of the risk factors. Then, mathematically, the CoVaR, denoted as ${\rm CoVaR}_{\alpha,\beta}$, satisfies the following equation
\begin{equation}\label{eqn:covar1}
    \Pr\{\pi(Z)\le {\rm CoVaR}_{\alpha,\beta} | \mu(Z)=q_\alpha\}=\beta,
\end{equation}
where $q_\alpha$ denotes the $\alpha$-VaR of the portfolio loss $\mu(Z)$.

When measuring systemic risks, the portfolios in consideration are typically at the institution level. They inevitably include many financial derivatives underlying the risk factors. Many of these derivatives lack of closed-form pricing formulas and, therefore, need to be priced by running additional simulation experiments. To simplify notation, we group all financial instruments in a portfolio together, and let $\mu(z)=\oE[X|Z=z]$ and $\pi(z)=\oE[Y|Z=z]$, where $X$ and $Y$ denote the aggregated discounted random losses of the two portfolios respectively at a future time $T$, by which all derivatives realize their payoffs, and the expectation is taken with respect to the risk-neutral measure \citep{hull2016options}. Then, we may rewrite the mathematical definition of CoVaR to incorporate the repricing feature, i.e., the CoVaR$_{\alpha,\beta}$ satisfies
\begin{equation}\label{eqn:covar2}
        \Pr\left\{\oE[Y|Z]\le {\rm CoVaR}_{\alpha,\beta}\, | \,\oE[X|Z]=q_\alpha\right\}=\beta.
\end{equation}
This paper aims to develop efficient Monte-Carlo estimators of CoVaR, where both $\mu(z)$ and $\pi(z)$ need to be estimated. To the best of our knowledge, this important problem has not been addressed in the literature.

Previous work by \cite{huang2022monte} addressed Monte Carlo estimation of CoVaR under the assumption that the closed-form expressions of $\mu(Z)$ and $\pi(Z)$ were available. They identified that, compared to single-portfolio risk measures like VaR, the key challenge in estimating CoVaR lies in handling the zero-probability event (ZPE), $\{\mu(Z) = q_\alpha\}$, which cannot be directly observed in the simulated data. To address this, \cite{huang2022monte}  proposed a batching estimation method, where the data is divided into batches, and each batch is used to generate an observation from the conditional distribution where the ZPE approximately holds. Their work shows that the convergence rate of this batching estimator is $\cO_{\rm P}(n^{-1/3})$, where $n$ is the sample size of the risk factors $Z$. This rate is slower than the typical convergence rate of $\cO_{\rm P}(n^{-1/2})$ for single-portfolio risk measure estimators, highlighting the added difficulty imposed by the ZPE in CoVaR estimation. 

When closed-form expressions of $\mu(z)$ and $\pi(z)$ are unavailable, they must be repriced using two-level simulations. In this setting, the outer-level simulation generates observations of the risk factors $Z$, while the inner-level simulation generates observations of the portfolio losses $X$ and $Y$, conditioned on the outer-level $Z$ observations. This two-level process adds considerable complexity to CoVaR estimation, as the generation of inner-level observations is far more computationally burdensome than generating outer-level scenarios.\footnote{We denote the time point at which we need the risk measure as $t=0$. Suppose that we are interested in the risk measure associated with a portfolio at a future time $\tau$, wherein the portfolio contains instruments with the longest maturity being $T$. We define $\tau$ as the risk horizon and typically we have $T\gg\tau$. The outer-level simulation entails advancing the risk factors from time $0$ to $\tau$. The inner-level simulation not only requires advancing the risk factors from $\tau$ to $T$—a significantly extended duration—but also necessitates the evaluation of payoffs across numerous derivatives.} Thus, the challenge posed by CoVaR estimation becomes much greater than the problem tackled by \cite{huang2022monte}, where closed-form expressions were assumed. In this paper, we evaluate the sampling efficiency of a CoVaR estimator in terms of its convergence rate with respect to $\Gamma$, which represents the total number of inner-level observations used to compute the estimator. Since generating outer-level scenarios is relatively simple, especially for large-scale portfolios common in CoVaR estimation, we overlook the sampling effort required for the outer-level simulation and focus primarily on the inner-level computational burden.

The repricing issue has also been addressed in the context of single-portfolio risk measurement, where the problem is often framed as estimating $\oE[g(\oE[X|Z])]$ for some nonlinear function $g(\cdot)$. A brief literature review of this topic sheds light on the available approaches and potential challenges. The main focus of the literature has been on a crucial question: {\it How many inner-level observations should be allocated to each outer-level scenario?} There are two primary approaches to addressing this question. The first is nested simulation, introduced by \cite{gordy2010nested}, which estimates $\mu(z) = \oE[X|Z = z]$ by averaging inner-level observations conditional on $Z = z$. They demonstrate that for $n$ outer-level scenarios of $Z$, the optimal allocation is to assign an order of $\sqrt{n}$ inner-level observations to each outer-level scenario. Under this allocation strategy, the estimator achieves a convergence rate of $\cO(\Gamma^{-1/3})$, where $\Gamma$ denotes the total number of inner-level observations. Importantly, this convergence rate is independent of the dimensionality of the risk factors, making the method robust against the curse of dimensionality.

The second approach is smoothing, developed by \cite{liu2010stochastic}, \cite{broadie2015risk}, \cite{hong2017kernel}, and \cite{wang2022smooth}, among others. The central idea behind this method is to incorporate information from neighboring outer-level scenarios to improve the accuracy of the point estimate $\oE[X|Z=z]$, allowing each outer-level scenario to require only one inner-level observation.\footnote{In practice, a constant number of inner-level observations, rather than just one, are often used, but this does not affect the convergence rate.} The convergence rate of the estimator depends on the smoothing technique employed, the dimensionality of the risk factors, and the smoothness of the underlying function $\mu(z)$. For instance, \cite{broadie2015risk} applied linear regression, achieving a convergence rate of $\cO_{\rm P}(\Gamma^{-1/2})$. However, this method relies on the careful selection of appropriate basis functions, and if the basis functions are not well-chosen, it may introduce an irreducible bias. On the other hand, \cite{hong2017kernel} used kernel smoothing, which achieves a convergence rate of $\cO(\Gamma^{-\min\{1/2, 2/(d+2)}\})$ in terms of RMSE, where $d$ is the dimensionality of the risk factors. Kernel smoothing suffers from the curse of dimensionality and performs poorly when $d$ is large. More recently, \cite{wang2022smooth} adopted kernel ridge regression (KRR), which can achieve a convergence rate of approximately $\cO_{\rm P}(\Gamma^{-\frac{2\nu-d}{4\nu+2d}})$, where $\nu$ represents the smoothness of the function $\mu(z)$. If $\mu(z)$ is sufficiently smooth, the convergence rate can approach approximately $\cO_{\rm P}(\Gamma^{-1/2})$, making KRR more robust in higher-dimensional settings than other smoothing methods.

In the CoVaR estimation, as considered in this paper, we face two simultaneous challenges: ZPE handling and repricing. For handling the ZPE, the batching estimator with $n$ outer-level scenarios has a convergence rate of $n^{-1/3}$. Regardless of whether the nested simulation or smoothing approach is applied for repricing, this results in the CoVaR estimator exhibiting low sampling efficiency. When using the nested simulation approach for repricing, we show that the optimal allocation strategy is to assign an order of $n^{1/3}$ inner-level observations to each of the $n$ outer-level scenarios. This configuration achieves an optimal convergence rate of $\cO_{\rm P}(\Gamma^{-1/4})$. On the other hand, when using the smoothing approach for repricing, each outer-level scenario is assigned only a single inner-level observation. In this case, the best possible convergence rate for the CoVaR estimator is $\cO_{\rm P}(\Gamma^{-1/3})$. Both approaches reveal low sampling efficiency, requiring significant computational effort to reach a desired level of precision. Therefore, a natural question arises: ``{\it Is it possible for a CoVaR estimator to achieve a convergence rate of $\cO_{\rm P}(\Gamma^{-1/2})$?}" Notice that, under the batching estimation scheme, the only possible sample allocation strategy to achieve a positive answer to this question is to allocate, on average, fewer than one inner-level observation to each outer-level scenario.

It turns out that achieving this is indeed possible! Before presenting our approach, let us begin with a thought experiment. Suppose we have approximated closed-form expressions for $\mu(z)=\oE[X|Z=z]$ and $\pi(z)=\oE[Y|Z=z]$ available in advance. In this case, we revert to the CoVaR estimation with closed-form expressions as described by \cite{huang2022monte}, eliminating the need for inner-level simulations to reprice the portfolios. Therefore, the key is to construct approximated closed-form expressions for $\mu(z)$ and $\pi(z)$ in the first stage, which can then be utilized to estimate CoVaR in the second stage. This approach allows the inner-level observations (along with their outer-level counterparts) to be used solely for constructing the surfaces in the first stage, while additional outer-level scenarios are employed to estimate CoVaR in the second stage, thereby allowing fewer inner-level observations than outer-level scenarios.

We call this two-stage approach ``the decoupled approach" to emphasize its critical characteristics of separating surface fitting from CoVaR estimation. While smoothing techniques, such as linear regression, kernel smoothing or KRR, are also used to fit the surfaces in the first stage, it is important to notice that the objective is different. In the original smoothing approach, the goal is to obtain accurate point estimates of $\mu(z)$ and $\pi(z)$ only for the $z$ values observed at the outer level. In contrast, the decoupled approach aims to construct functional approximations of $\mu(z)$ and $\pi(z)$ for all $z$ values in its domain.

The decoupled approach not only enables the development of efficient CoVaR estimators but also simplifies the asymptotic analysis of these estimators. In alignment with the two stages of the decoupled approach, the error of the CoVaR estimator can be split into two components: the functional-approximation-related error and the batching-estimation-related error. While the latter is relatively straightforward to handle and has minimal impact (as it only involves outer-level scenarios), managing the former requires innovative methods. We develop a functional approach to analyze this error, demonstrating that the convergence rate of the decoupled CoVaR estimator is the same as the convergence rate of approximating $\mu(z)$ and $\pi(z)$, measured by their $L^\infty$ norms over the entire domain. The reason for focusing on $L^\infty$ norms, rather than more commonly used $L^2$ norms, is due to the ZPE, as $L^2$ convergence does not guarantee convergence at ZPEs. Notice that this finding is independent of the specific smoothing technique employed for functional approximations. Therefore, it can be seen as a general plug-and-play framework that works with any smoothing technique, as long as the $L^\infty$ convergence rates can be derived.

The convergence rate result indicates that for the decoupled CoVaR estimator to achieve good sampling efficiency, we must employ smoothing techniques with fast $L^\infty$ convergence rates when approximating $\mu(z)$ and $\pi(z)$.  First, because CoVaR addresses institutional-level portfolios, the dimensionality of the risk factors is usually high. Thus, we need smoothing techniques that can handle high-dimensional data without succumbing to the curse of dimensionality. While most smoothing techniques are evaluated in terms of $L^2$ convergence rates, we specifically require $L^\infty$ convergence. Interestingly, both issues are ultimately related to the smoothness of the portfolio loss functions $\mu(z)$ and $\pi(z)$. In the literature, \cite{wang2022smooth} have already demonstrated that KRR achieves excellent $L^2$ convergence rates when the portfolio loss functions are sufficiently smooth. We find that neural networks exhibit similar performance. Moreover, by leveraging the Gagliardo-Nirenberg interpolation inequality \citep{nirenberg1959elliptic}, we show that $L^\infty$ and $L^2$ convergence rates are approximately equivalent, provided that the loss functions are smooth enough.  This brings us to another key question: {\it ``Are the portfolio loss functions $\mu(z)$ and $\pi(z)$ sufficiently smooth?"} It is important to note that \cite{wang2022smooth} made this assumption without offering formal theoretical justification. In this paper, we provide proof that under very general conditions, these loss functions are indeed infinitely differentiable. With this key finding, we demonstrate that the decoupled CoVaR estimator can achieve a convergence rate of approximately $\cO_{\rm P}(\Gamma^{-1/2})$, provided that smoothing techniques like linear regression (with appropriate basis functions), KRR (with appropriate kernel functions), or neural networks (with appropriate activation functions) are employed.

The rest of the paper is organized as follows: Section \ref{sec:BE} introduces the batching estimator from \cite{huang2022monte} and presents two na\"ive estimators. Section \ref{sec:decoupling} provides a detailed explanation of the decoupled approach and its convergence rate analysis. In Section \ref{sec:smoothing}, we discuss the smoothness of the loss functions and the corresponding smoothing techniques. Numerical results are presented in Section \ref{sec:num}, with conclusions offered in Section \ref{sec:conclusions}. The proofs and relevant technical conditions are provided in the Appendix.

\section{Batching Estimator and Two Na\"{i}ve Nested Estimators}\label{sec:BE}

In this section, we first briefly review the batching estimator of CoVaR, proposed by \cite{huang2022monte} where the closed-form expressions of the loss functions $\mu(z)$ and $\pi(z)$ are available. This estimator serves as the foundation of the estimators proposed in this paper, which are designed to handle the repricing issue that is ubiquitous in practice but not considered by \cite{huang2022monte}. We then present two na\"ive estimators of CoVaR that directly incorporate the nested simulation approach and the smoothing approach, two commonly used repricing approaches for single-portfolio risk measurement, into the batching estimator. We analyze the convergence rates of these two estimators, and explain why they are inefficient in solving the CoVaR estimation problem.

\subsection{The Batching Estimator}\label{subsec:BE}

When the closed-form expressions of $\mu(z)$ and $\pi(z)$ are available, the critical issue of CoVaR estimation is the handling of the ZPE $\{\mu(Z)=q_\alpha\}$, where $q_\alpha$ is the $\alpha$-VaR of $\mu(Z)$. \cite{huang2022monte} proposed a batching estimation method to address this issue, see Algorithm \ref{alg:BE} for the detailed estimation algorithm. 

\begin{algorithm}[h]
\caption{Batching Estimation of CoVaR}\label{alg:BE}
\begin{description}
\item[\bf Step 1.] Generate $n$ scenarios of risk factors $\{Z_i\}_{i=1}^n$ and compute the corresponding observations of two portfolio losses $\{(\mu(Z_i), \pi(Z_i))\}_{i=1}^n$. For convenience, denote them as $\{(\mu_i,\pi_i)\}_{i=1}^n$.	\smallskip
	
\item[\bf Step 2.] Divide the data $\{(\mu_i,\pi_i)\}_{i=1}^n$ into $k$ batches and each batch has $h$ observations with $n=k\times h$, and denote the observations in the $i$-th batch as $\{(\mu_{i,j}, \pi_{i,j})\}_{j=1}^h$, $i=1,2,\ldots,k$. 
	\smallskip
	
\item[\bf Step 3.] For each batch (say $i$-th batch), sort $\mu_{i,1},\ldots,\mu_{i,h}$ from lowest to highest, denoted by $\mu_{i,(1)}\leq \mu_{i,(2)}\leq \cdots \leq \mu_{i,(h)}$, where $\mu_{i,(j)}$ denotes the $j$-th smallest value. Let $\check \pi_i = \pi_{i,(\lceil{\alpha h}\rceil)}$, where $\pi_{i,(\lceil{\alpha h}\rceil)}$ is the concomitant of $\mu_{i,(\lceil{\alpha h}\rceil)}$, i.e., $\pi_{i,(\lceil{\alpha h}\rceil)}$ shares the same realization of risk factors with $\mu_{i,(\lceil{\alpha h}\rceil)}$.
	\smallskip
	
\item[\bf Step 4.] Sort $\check \pi_1,\ldots,\check \pi_k$ from lowest to highest, denoted by $\check \pi_{(1)}\le\check \pi_{(2)}\le\cdots\leq\check \pi_{(k)}$. Then, $\check \pi_{(\lceil{\beta k}\rceil)}$ is the batching estimator of ${\rm CoVaR}_{\alpha,\beta}$.
\end{description}
\end{algorithm}

The batching estimation method first divides the simulation observations $\{(\mu_i,\pi_i)\}_{i=1}^n$ into $k$ batches and each batch has $h$ observations. Let $\{(\mu_{i,j}, \pi_{i,j})\}_{j=1}^h$ denote the observations of the $i$-th batch, $i=1,2,\ldots,k$. Then, for each batch (say $i$), it uses $\mu_{i,1},\ldots,\mu_{i,h}$ to find the $\lceil h\alpha\rceil$-th order statistic $\mu_{i,(\lceil h\alpha\rceil)}$, which is a strongly consistent estimator of $q_\alpha$ as $h\to\infty$ \citep{serfling2009approximation}. Then, its corresponding concomitant $\pi_{i,(\lceil h\alpha\rceil)}$, denoted by $\check\pi_i$, is an observation of $\pi(Z)$ that approximately satisfies the ZPE $\{\mu(Z)=q_\alpha\}$. Since each batch generates one such observation, $k$ batches give an independent and identically distributed sample, denoted as $\check \pi_1,\ldots,\check \pi_k$, which may then be used to estimate its $\beta$-VaR, which is the batching estimator of ${\rm CoVaR}_{\alpha,\beta}$.

\cite{huang2022monte} proved that the batching estimator is strongly consistent as $n\to\infty$. It can achieve the best rate of convergence rate $\cO_{\rm P}(n^{-1/3})$ by letting $\sqrt{k}/h\rightarrow c$, for some constant $c>0$. Notice that a standard VaR estimator has a rate of convergence of $\cO_{\rm P}(n^{-1/2})$ \citep{serfling2009approximation}. Therefore, estimating CoVaR is significantly more difficult and may require a substantially larger sample size to achieve the desirable level of precision than estimating VaR.

\subsection{Two Na\"ive Nested Estimators}
\label{subsec:naive}
Now we consider the nested CoVaR estimation problem where the closed-form expressions of $\mu(z)$ and $\pi(z)$ are not available, and one has to use inner-level observations of $X$ and $Y$ conditioned on $Z=z$ to estimate them. Furthermore, we keep the framework of the batching estimation to handle the ZPE, as it is convenient and easy to understand.

Suppose that we generate $l$ inner-level observations $(X_{ij},Y_{ij})_{j=1}^l$ for each outer-level scenario $Z_i$, $i=1,\ldots,n$. Depending on whether the nested simulation approach or the smoothing approach is used $l$ may be increasing in $n$ or may be a constant integer (such as one). Let $\bar{X}_i={1 \over l}\sum_{j=1}^l X_{ij}$ and $\bar{Y}_i={1\over l}\sum_{j=1}^l Y_{ij}$ for all $i=1,\ldots,n$. The nested simulation approach uses $\bar X_i$ and $\bar Y_i$ as estimates of $\mu(Z_i)$ and $\pi(Z_i)$, while the smoothing approach uses a smoothing technique on $\{(\bar X_i,\bar Y_i)\}_{i=1}^n$ to obtain estimates of $\mu(Z_i)$ and $\pi(Z_i)$. In both cases, we denote the estimates as $\{(\hat\mu_i,\hat\pi_i)\}_{i=1}^n$. Then, we may plug them into Algorithm \ref{alg:BE} to obtain two na\"ive nested estimators of CoVaR (see Algorithm \ref{alg:BE2}).

\begin{algorithm}[h]
\caption{Na\"ive Nested Estimation of CoVaR}\label{alg:BE2}
\begin{description}
\item[\bf Step 1.] Generate $n$ scenarios of risk factors $\{Z_i\}_{i=1}^n$. For each $Z_i$ $i=1,\ldots,n$, generated $l$ observations of $(X,Y)|Z=Z_i$, denoted by $(X_{ij},Y_{ij})_{j=1}^l$, and calculate $\bar X_i$ and $\bar Y_i$. Use either the nested simulation approach or the smoothing approach with a chosen smoothing technique to obtain estimates of $\{\mu(Z_i),\pi(Z_i)\}_{i=1}^n$, denoted by $\{(\hat\mu_i,\hat\pi_i)\}_{i=1}^n$.
\smallskip
	
\item[\bf Step 2.] Divide the data $\{(\hat\mu_i,\hat\pi_i)\}_{i=1}^n$ into $k$ batches and each batch has $h$ observations with $n=k\times h$, and denote the observations in the $i$-th batch as $\{(\hat\mu_{i,j}, \hat\pi_{i,j})\}_{j=1}^h$, $i=1,2,\ldots,k$. 
	\smallskip
	
\item[\bf Step 3.] For each batch (say $i$-th batch), sort $\hat\mu_{i,1},\ldots,\hat\mu_{i,h}$ from lowest to highest, denoted by $\hat\mu_{i,(1)}\leq \hat\mu_{i,(2)}\leq \cdots \leq \hat\mu_{i,(h)}$, where $\hat\mu_{i,(j)}$ denotes the $j$-th smallest value. Let $\check {\hat\pi}_i = \hat\pi_{i,(\lceil{\alpha h}\rceil)}$, where $\hat\pi_{i,(\lceil{\alpha h}\rceil)}$ is the concomitant of $\hat\mu_{i,(\lceil{\alpha h}\rceil)}$, i.e., $\hat\pi_{i,(\lceil{\alpha h}\rceil)}$ shares the same realization of risk factors with $\hat\mu_{i,(\lceil{\alpha h}\rceil)}$.
	\smallskip
	
\item[\bf Step 4.] Sort $\check {\hat\pi}_1,\ldots,\check{\hat\pi}_k$ from lowest to highest, denoted by $\check{\hat\pi}_{(1)}\le\check {\hat\pi}_{(2)}\le\cdots\leq\check{\hat\pi}_{(k)}$. Then, $\check{\hat\pi}_{(\lceil{\beta k}\rceil)}$ is the nested estimator of ${\rm CoVaR}_{\alpha,\beta}$.
\end{description}
\end{algorithm}

Algorithm \ref{alg:BE2} is almost identical to Algorithm \ref{alg:BE}, except that we altered Step 1 to generate $\{(\hat\mu_i,\hat\pi_i)\}_{i=1}^n$ to approximate $\{(\mu_i,\pi_i)\}_{i=1}^n$ in Algorithm \ref{alg:BE}. While Algorithm \ref{alg:BE2} is straight-forward and easy to understand, their estimators are not necessarily efficient. In the rest of this section, we analyze the convergence rates of the two estimators of Algorithm \ref{alg:BE2}, one using the nested simulation approach to obtain  $\{(\hat\mu_i,\hat\pi_i)\}_{i=1}^n$, while the other use a smoothing technique.

To compare the efficiency of different estimators, we need a common yardstick. In this paper, we follow the tradition of the literature and consider the convergence rate with respect to the total sampling effort \citep{gordy2010nested,hong2017kernel,wang2022smooth}. As the sample includes both inner-level observations and outer-level scenarios, we let the cost of generating an inner-level observation be $1$ and the cost of generating an outer-level scenario be $\gamma$. As the inner-level observation involves much longer time horizon and a large number of derivatives, its generation is significantly more difficult than the outer-level scenario. Therefore, $\gamma$ is typically much smaller than 1 and is often negligible (i.e., $\gamma$ may be treated as 0). In this paper, we use the total number of inner-level observations, denoted by $\Gamma$, to measure the sampling effort, and compare the convergence rates of different estimators by letting $\Gamma\to\infty$.\footnote{We want to emphasize that, even though the comparison based on sampling cost is widely used in the literature of nested estimation, it overlooks the computational overhead that may be quite significant, especially for some smoothing techniques that require tuning when fitting the surfaces (e.g., the kernel method, KRR and neural networks) and need significant computation to evaluate (e.g., the kernel method, kriging and KRR).}

First, we consider the estimator with the nested simulation approach for repricing, where only the sample means $\bar X_i$ and $\bar Y_i$ are used to estimate $\mu(Z_i)$ and $\pi(Z_i)$. We have the following proposition on the convergence rate of the root mean squared error (RMSE), where $k$ is the number of batches, $h$ is the batch size and $l$ is the number of inner-level observations for each outer-level scenario. Notice that $k\times h=n$, which is the total number of outer-level scenarios. The proof sketch of this proposition can be found in Appendix~\ref{sec:other-proofs}.

\begin{proposition}\label{prop:CoVaR_SNS}
The RMSE of the CoVaR estimator based on the nested simulation approach is of order $\cO_{\rm P}(k^{-1/2}+h^{-1}+l^{-1})$, which implies that the optimal rate of convergence is $\cO_{\rm P}(\Gamma^{-1/4})$ when $l=c_1\Gamma^{1/4}$, $h=c_2\Gamma^{1/4}$ and $k={1 \over c_1c_2}\Gamma^{1/2}$ for some constants $c_1, c_2>0$.
\end{proposition}

Notice the nested estimation of CoVaR is in general more difficult than the estimation of CoVaR without repricing or the estimation of VaR with repricing. From the literature, we know that the convergence rates of the latter two are both $\cO_{\rm P}(\Gamma^{-1/3})$ \citep{huang2022monte,gordy2010nested}. Therefore, it is not surprising that the convergence rate of the nested-simulation based CoVaR estimator demonstrated in Proposition \ref{prop:CoVaR_SNS} is worse than $\cO_{\rm P}(\Gamma^{-1/3})$. This convergence rate is certainly not desirable. It implies that a very large amount of sampling effort may be necessary to achieve an acceptable level of precision. Given that the inner-level observations are very costly to generate, this estimator is in general not recommended.

Second, we consider the estimator that uses the smoothing approach for repricing, where all sample means $(\bar X_i,\bar Y_i)_{i=1}^n$ are used to estimate $\mu(Z_i)$ and $\pi(Z_i)$ through a smoothing technique. Analyzing the precise convergence rate of this estimator is more challenging because the smoothing approach introduces dependence across all $(\hat\mu_i,\hat\pi_i)_{i=1}^n$, linking the sample means and complicating the analysis. However, we can provide a heuristic argument that the convergence rate is unlikely to exceed $\cO_{\rm P}(\Gamma^{-1/3})$. Note that $l \geq 1$ in the estimator, meaning that even if the approximation error of $(\hat\mu_i,\hat\pi_i)_{i=1}^n$ is negligible, i.e., we treat it as $(\mu_i,\pi_i)_{i=1}^n$, the estimator cannot converge faster than the batching estimator with closed-form expressions for $\mu(z)$ and $\pi(z)$. Therefore, the convergence rate is upper bounded by $\cO_{\rm P}(\Gamma^{-1/3})$.

Even if this $\cO_{\rm P}(\Gamma^{-1/3})$ convergence rate is achievable, it is not particularly efficient. The high computational cost of generating inner-level observations motivates us to explore whether the convergence rate can be further improved. To achieve this, the key is to reduce the sample size of the inner-level observations relative to the outer-level scenarios, i.e., allocating, on average, fewer than one inner-level observation per outer-level scenario. This represents a clear departure from traditional sample allocation strategies and is not feasible within the na\"ive estimation framework presented in Algorithm \ref{alg:BE2}. In the next section, we will introduce the decoupled approach, which makes this possible.

\section{The Decoupled Approach}\label{sec:decoupling}

Recall that there are two main challenges in estimating CoVaR: the ZPE and repricing. These two challenges are closely intertwined: the repriced portfolio losses are necessary to identify the outer-level scenarios where the ZPE holds approximately. The na\"ive estimators introduced in Section \ref{sec:BE} couple these challenges together by repricing portfolio losses for every outer-level scenario. Because a large number of outer-level scenarios are needed to identify those where the ZPE holds approximately, the na\"ive estimators also require a correspondingly large number of inner-level observations for repricing through either the nested simulation approach or the smoothing approach.

In this section, we introduce a flexible algorithmic framework that decouples these two challenges, addressing them one at a time. This framework allows for the use of different smoothing techniques in the repricing stage. We then show that the convergence rate of the new decoupled estimator may be derived through a decoupled mindset as well, and the convergence rate ultimately depends on the convergence property of the chosen smoothing technique, which will be further discussed in Section \ref{sec:smoothing}.

\subsection{The Algorithm}

The decoupled approach operates in two distinct stages. In the first stage, it simulates $m$ outer-level scenarios and $l$ inner-level observations conditioned on each outer-level scenario (where $l$ can be as small as $1$), and uses these observations to construct function approximations to the portfolio loss functions $\mu(z)$ and $\pi(z)$ using a smoothing technique. In the second stage, it generates $n$ new outer-level scenarios and employs the function approximations from the first stage as to execute the batching estimation algorithm to obtain the CoVaR estimator. We summarize it in Algorithm \ref{alg:CoVaR}.

\begin{algorithm}[t]\label{alg:CoVaR}
\caption{The Decoupled Approach of CoVaR Estimation}\label{alg:CoVaR}
\begin{algorithmic}
\State \textbf{The First Stage: Two-level Simulation and Function Approximation}
\IndState $-$ Generate $m$ independent outer-level scenarios $\{\Tilde{Z}_{i}\}_{i=1}^m$.
\IndState $-$ For each scenario $\tilde Z_{i}, i=1,\cdots, m$, generate $l$ independent outcomes of $X$ and $Y$, denoted by $\{\tilde X_{ij}\}_{j=1}^l$ and $\{\tilde Y_{ij}\}_{j=1}^l$.  Compute $\tilde{X}_i = \frac{1}{l}\sum_{j=1}^l\tilde X_{ij}$ and $\tilde{Y}_i = \frac{1}{l}\sum_{j=1}^l\tilde Y_{ij}$.
\IndState $-$ Fit the functions of portfolio losses $\mu(\cdot)$ and $\pi(\cdot)$ using $\{(\Tilde{Z}_{i},\Tilde{X}_{i})\}_{i=1}^{m}$ and $\{(\Tilde{Z}_{i},\Tilde{Y}_{i})\}_{i=1}^{m}$ by a selected smoothing technique. Let $\tilde{\mu}_m(\cdot)$ and $\tilde{\pi}_m(\cdot)$ denote the approximated functions of the portfolio losses.
\State \textbf{The Second Stage: Batching Estimation}
\IndState $-$ Generate $n$ independent outer-level scenarios $\{Z_{i}\}_{i=1}^n$.
\IndState $-$ Compute the corresponding $\tilde{\mu}_m(Z_i)$ and $\tilde{\pi}_m(Z_i)$ and denote them as $\hat{\mu}_i$ and $\hat{\pi}_i$ for $i=1,\ldots, n$.
\IndState $-$ Divide the data into $k$ batches with $h$ samples in each batch and denote the observations in the $s$-th batch as $\{\hat{\mu}_{s,t}\}_{t=1}^{h}, s=1,\ldots,k$.
\IndState $-$ For each batch (say $s$-th batch), sort $\{\hat{\mu}_{s,t}\}_{t=1}^{h}$ from lowest to highest, denoted by $\hat{\mu}_{s,(1)}\leq\ldots\leq\hat{\mu}_{s,(h)}$. 
\IndState $-$ Select the $\hat{\mu}_{s,(\lceil h\alpha \rceil)}, s=1,\ldots,k$ as the estimator of  $\alpha$-th quantile of $\hat{\mu}(Z)$. Let $\check{\pi}_s = \hat{\pi}_{s,(\lceil h\alpha \rceil)}$. 
\IndState $-$ Sort $\check{\pi}_s$ from smallest to largest and estimate CoVaR by $\hat{\theta}^{NB}_{m,n} =\check{\pi}_{(\lceil k\beta \rceil)}$.
\end{algorithmic}
\end{algorithm}

In the decoupled approach, the repricing and ZPE challenges are addressed separately in the first and second stages, respectively. Through this decoupling, the large number of outer-level scenarios necessary for the second stage no longer demand corresponding inner-level observations, which allows for far more outer-level scenarios than inner-level ones.

Another way to understand the decoupled approach is by revisiting the original batching estimation in Algorithm \ref{alg:BE}, where closed-form expressions of the portfolio loss functions $\mu(z)$ and $\pi(z)$ are known. Since these expressions are unknown in our context, Algorithm \ref{alg:CoVaR} introduces a separate first stage to approximate them. Although this interpretation may seem straightforward, it obscures the crucial reasoning behind the decoupled strategy described earlier. It is worth noting that the decoupled approach can also apply to single-portfolio risk measure estimation, though it wasn't explicitly used by \cite{hong2017kernel} or \cite{wang2022smooth}. It only becomes evident in CoVaR estimation that decoupled approach can offer substantial benefits.

\subsection{Analysis of the Convergence Rate}\label{subsec:analysis}

Unlike the analysis in the literature on single-portfolio risk measurement, which is often intertwined with the specific smoothing techniques used, the goal of this subsection is to establish a general result on the convergence rate of the decoupled CoVaR estimator that accommodates different smoothing techniques. The key ingredient for achieving this is the decoupled structure, which allows us to separate the total estimation error into two parts: the function-approximation-related error (FA error) and the batching-estimation-related error (BE error), corresponding to the two stages of the approach. The BE error analysis follows directly from the results of \cite{huang2022monte} and is relatively straightforward. However, the FA error analysis is more complex and requires significant innovation. In this subsection, we adopt a functional perspective and connect the FA error to the $L^\infty$ norm of the functional approximation error.

\subsubsection{Notation and Preliminary Results.}

Let $q$ denote the $\alpha$-VaR of $\mu(Z)$ and $\theta$ denote the $(\alpha,\beta)$-CoVaR of $(\mu(Z),\pi(Z))$ for simplicity. Then, $q$ and $\theta$ satisfy $\oPr\{\mu(Z)\le q\}=\alpha$ and $\oPr\{\pi(Z)\le \theta|\mu(Z)=q\}=\beta$, respectively. Notice that $\theta$ is the CoVaR that we want to estimate. Conditioned on the fitted portfolio loss functions $\mum(\cdot)$ and $\pim(\cdot)$, we let $\qm$ denote the $\alpha$-VaR of $\mum(Z)$ and $\tm$ denote the $(\alpha,\beta)$-CoVaR of $(\mum(Z),\pim(Z))$. Then, $\qm$ and $\tm$ satisfy $\oPr_m\{\mum(Z)\le \qm\}=\alpha$ and $\oPr_m\{\pim(Z)\le \tm|\mum(Z)=\qm\}=\beta$, respectively, where $\oPr_m$ denotes the probability conditioned on $\mum(\cdot)$ and $\pim(\cdot)$. Therefore, $\qm$ and $\tm$ are both random variables whose values depend on the first-stage inner-level and outer-level scenarios that are used to construct $\mum$ and $\pim$. Let $\tmn$ denote the CoVaR estimator produced by Algorithm \ref{alg:CoVaR}. In this section we are interested in finding the convergence rate of $\tmn$ to $\theta$.

Let functions $F_\mu$ and $f_\mu$ denote the distribution function and density function of $\mu(Z)$, i.e., $F_\mu(x)=\oPr\{\mu(Z)\le x\}$ and $f_\mu(x)=F'_\mu(x)$, and let functions $F_{\mum}$ and $f_{\mum}$ denote the conditional distribution function and conditional density function of $\mum(Z)$ conditioned on $\mum(\cdot)$, i.e., $F_{\mum}(x)=\oPr_m\{\mum(Z)\le x\}$ and $f_{\mum}(x)={d\over dx}F_{\mum}(x)$. In the similar manner, we let $G_{\mu,\pi}(x,y)=\oPr\{\pi(Z)\le y|\mu(Z)=x\}$ and $G_{\mum,\pim}(x,y)=\oPr_m\{\pim(Z)\le y|\mum(Z)=x\}$.

We may also treat $F_\mu(x)$, $f_\mu(x)$ and $G_{\mu,\pi}(x,y)$ as functionals of $\mu$ and $\pi$. For fixed values of $x,y$, we can define functionals of $\mu$ and $\pi$ by $\cF_x[\mu] = F_\mu(x)$ and $\cG_{x,y}[\mu,\pi] = G_{\mu,\pi}(x,y)$. Then, we have $\cF_x[\mum] = F_{\mum}(x)$ and $\cG_{x,y}[\mum,\pim] = G_{\mum,\pim}(x,y)$. Taking the viewpoint of functionals is an important innovation of our approach. It allows us to link the estimation error of CoVaR directly to the estimation errors of the functions $\mu(\cdot)$ and $\pi(\cdot)$, and it is more inline with the decoupled algorithm.

Besides, the $L^p$ norm will be used frequently through this paper, which is defined as
\begin{eqnarray*}
&\|f\|_{L^p}=\left(\int_{\Omega}|f(z)|^p {\rm d}P(z)\right)^{1/p} &\quad \text{ if } p<+\infty,\\
&\|f\|_{L^\infty}= {\rm ess}\sup_{z\in \Omega}|f(z)| &\quad \text{ if } p=+\infty,
\end{eqnarray*}
where $\Omega$ is the domain of $z$ and $P$ is the distribution of $z$. If $\|f\|_{L^p}<\infty$, then we say $f\in L^p$. For convenience, we abbreviate $\|\cdot\|_{L^p}$ by $\|\cdot\|_{p}$ in this paper.

To analyze the convergence rate of the CoVaR estimator, we need the following assumption on the convergence rate of the fitted surfaces $\mum(\cdot)$ and $\pim(\cdot)$.
\begin{assumption}\label{asmp:err_rate}
For sufficiently large $m$, $\|\mum-\mu\|_\infty = \cO_{\oP}(a_m)$ and $\|\pim-\pi\|_\infty = \cO_{\oP}(a_m)$, where $\{a_m\}$ is a deterministic sequence with $\lim_{m\rightarrow\infty} a_m = 0$.
\end{assumption}
Notice that Assumption~\ref{asmp:err_rate} is a general assumption that allows many different types of smoothing techniques. Our goal is to establish a general theoretical framework for the decoupled algorithm and consider the choice of particular smoothing technique as a separate issue.

With the assumption, we can establish some preliminary results that are used later. These results all need additional assumptions on the smoothness of various involved functions. These conditions are tedious and difficult to verify. But they are generally satisfied and are used widely in the literature \citep{gordy2010nested, broadie2015risk, hong2017kernel}. To maintain readability, we use ``regularity conditions" to refer them and state them in detail in the Appendix. 

\begin{lemma}\label{lem:f}
Under Assumption \ref{asmp:err_rate} and certain regularity conditions, for any sequence $\{x_m\}$ such that $x_m\rightarrow x$ as $m\to\infty$, we have
 $f_{\mum}(x_m)\rightarrow f_\mu(x)$ and $f'_{\mum}(x_m)\rightarrow f'_{\mu}(x)$ almost surely as $m\to\infty$. 
\end{lemma}

We provide a proof sketch of Lemma \ref{lem:f}. The more rigorous proof along with the regularity conditions are provided in Appendix~\ref{sec:lemma_proof}. 
\begin{proof}{Proof sketch.}
    Let $\eta_m(Z)=a_m^{-1}[\mum(Z)-\mu(Z)]$. Assumption \ref{asmp:err_rate} suggests that $\eta_m(Z)$ has a non-trivial limiting distribution as $m\to\infty$. Let $b_m(u,v)$ denote the joint density of $(\mu(Z),\eta_m(Z))$ conditioned on $\mum(\cdot)$. Then,
\begin{eqnarray*}
    F_{\mum}(x)&=&\oPr_m\left\{\mum(Z)\le x\right\}=\oPr_m\left\{\mu(Z)+a_m\eta_m(Z)\le x\right\}=\int_\mathbb{R}\int_{-\infty}^{x-a_m v}b_m(u,v)dudv,\\
    F_{\mu}(x)&=&\oPr\left\{\mu(Z)\le x\right\}=\oPr_m\left\{\mu(Z)\le x\right\}=\int_\mathbb{R}\int_{-\infty}^{x}b_m(u,v)dudv.
\end{eqnarray*}
Differentiating with respect to $x$, we have $f_{\mum}(x)=\int_\mathbb{R} b_m(x-a_m v,v)dv$ and $f_{\mu}(x)=\int_\mathbb{R} b_m(x,v)dv$. Then, applying the mean value theorem, we have
\begin{eqnarray*}
    f_{\mum}(x_m)- f_\mu(x) &=&\int_\mathbb{R} \left[b_m(x_m-a_m v,v)-b_m(x,v)\right] dv \\
    &=& \int_\mathbb{R} {\partial\over\partial x}b_m(x^*,v)(x_m-a_mv-x) dv\quad\textnormal{for some $x^*\in[x_m-a_mv,x]$}\\
    &=& (x_m-x)\int_\mathbb{R} {\partial\over\partial x}b_m(x^*,v) dv + a_m \int_\mathbb{R} {\partial\over\partial x}b_m(x^*,v) vdv.
\end{eqnarray*}
As both $x_m-x$ and $a_m$ converge to zero as $m\to\infty$, we have $f_{\mum}(x_m)\to f_\mu(x)$ almost surely as long as $b_m(u,v)$ satisfies certain regularity conditions almost surely. Similarly, we can show that  $f'_{\mum}(x_m)\to f'_\mu(x)$. \hfill$\square$\\[-10pt] 
\end{proof}

\begin{lemma}\label{lem:F}
Under Assumption \ref{asmp:err_rate} and certain regularity conditions, 
\[ \cF_q[\mum]-\cF_q[\mu]= \cO_{\oP}(a_m).\]
\end{lemma}

The conclusion of Lemma \ref{lem:F} appears to be a direct consequence of a first-order Taylor expansion of $\cF_q[\mu]$ at $\mu$ as $\|\mum-\mu\|_\infty=\cO_{\oP}(a_m)$. However, there are two difficulties. First, there are different ways to define a functional derivative. Second, it is not clear why it focuses on the $L^\infty$-norm of $\mum-\mu$ instead of other types of norms. We provide a proof sketch of the lemma to address these two difficulties. The more rigorous proof along with the regularity conditions are provided in Appendix~\ref{sec:lemma_proof}.

\begin{proof}{Proof sketch.}
Inspired by the definition of Gateaux derivative for functionals \citep{serfling2009approximation,engel2011density}, we consider ${d\over dw}\cF_q[\mu+w\eta_m]$ where $\eta_m=a_m^{-1}(\mum-\mu)$ may be viewed as a direction in the functional space. Notice that $\cF_q[\mu+w\eta_m]=\oPr_m\{\mu(Z)+w\eta_m(Z)\le q\}$. Its derivative with respect to $w$ may be computed using the probability sensitivity formula (Theorem 1 of \citealt{hong2009estimating}). Therefore, we have
\[
    {d\over dw}\cF_q[\mu+w\eta_m]=-\phi_{\mu+w\eta_m}(q)\cdot \oE_m[\eta_m(Z)|\mu(Z)+w\eta_m(Z)=q],
\]
where $\phi_{\mu+w\eta_m}(\cdot)$ is the density of $\mu(Z)+w\eta_m(Z)$ conditioned on $\mum$.

Now we can apply the first-order Taylor expansion,
\begin{eqnarray*}
    \Big|\cF_q[\mum]-\cF_q[\mu] \Big| &=& \Big|\cF_q[\mu+a_m\eta_m]-\cF_q[\mu] \Big| =\Big|\phi_\mu(q)\cdot \oE_m[\eta_m(Z)|\mu(Z)=q]\cdot a_m\Big| + o(a_m) \\
    &=& \Big|\phi_\mu(q)\cdot \oE_m\left[\mum(Z)-\mu(Z)|\mu(Z)=q\right]\Big| + o(a_m) \\
    &\le & \phi_\mu(q)\cdot \|\mum-\mu\|_\infty + o(a_m).
\end{eqnarray*}
The last step shows why the $L^\infty$-norm of $\mum-\mu$ is needed. This is because the $\{\mu(Z)=q\}$ may be a ZPE and other types of norms, e.g., $L^2$-norm, cannot bound $\mum-\mu$ conditioned on $\mu(Z)=q$. As $ \|\mum-\mu\|_\infty=\cO_{\oP}(a_m)$, we have $\cF_q[\mum]-\cF_q[\mu]= \cO_{\oP}(a_m)$ as well.
\hfill$\square$\\[-10pt] 
\end{proof}

\begin{lemma}\label{lem:g}
    Under Assumption \ref{asmp:err_rate} and certain regularity conditions, for any sequence $\{(x_m,y_m)\}$ such that $(x_m,y_m)\rightarrow(x,y)$ as $m\to\infty$, we have
\[
 {\partial\over\partial x}G_{\mum,\pim}(x_m,y_m)\rightarrow {\partial\over\partial x}G_{\mu,\pi}(x,y)\ \ {\rm and}\ \ 
 {\partial\over\partial y}G_{\mum,\pim}(x_m,y_m)\rightarrow {\partial\over\partial y}G_{\mu,\pi}(x,y)
\]
almost surely as $m\to\infty$. 
\end{lemma}

\begin{lemma}\label{lem:G}
    Under Assumption \ref{asmp:err_rate} and certain regularity conditions, 
\[ \cG_{q,\theta}[\mum,\pim]-\cG_{q,\theta}[\mu,\pi] = \cO_{\oP}(a_m).\]
\end{lemma}

Lemmas \ref{lem:g} and $\ref{lem:G}$ are extensions of Lemmas \ref{lem:f} and \ref{lem:F}, extending from uni-variate functions $F_\mu(\cdot)$ and $\cF_q(\cdot)$ to bi-variate functions $G_{\mu,\pi}(\cdot,\cdot)$ and $\cG_{q,\theta}[\cdot,\cdot]$. The proofs of these two lemmas are similar to those of Lemmas \ref{lem:f} and \ref{lem:F}. We include them in Appendix~\ref{sec:lemma_proof} along with the required regularity conditions.

\subsubsection{Convergence Rate of the CoVaR Estimator.}

Notice that the estimation error of the CoVaR estimator $\tmn$ may be decomposed into two parts, i.e., $\tmn-\theta=(\tm-\theta)+(\tmn-\tm)$. The first part $(\tm-\theta)$ is the error caused by function approximations of $\mu$ and $\pi$ in the first stage of Algorithm \ref{alg:CoVaR}; and the second part $(\tmn-\tm)$  is the error caused by the batching estimation in the second stage of the algorithm. To analyze the rate of convergence of $\tmn$, we analyze separately these two errors, which we call the FA error and BE error respectively.

We first study the convergence rates of $\qm$ and $\tm$, which are the true VaR and CoVaR values once the surface approximations $\mum$ and $\pim$ are obtained in the first stage of Algorithm \ref{alg:CoVaR}. Their errors $\qm-q$ and $\tm-\theta$ are completely determined by the first stage sample, and $\tm-\theta$ is the FA error that we are interested in.

\begin{proposition}\label{prop:qm}
Under Assumption \ref{asmp:err_rate} and the regularity conditions of Lemmas \ref{lem:f} and \ref{lem:F}, $\qm-q= \cO_{\oP}(a_m)$.
\end{proposition}

\begin{proof}{Proof.}
By the mean value theorem, it can be shown that (see Lemma~\ref{lem:bi-taylor} in Appendix~\ref{sec:lemma_proof}), as $m\to \infty$, $F_{\mum}(\qm)=F_{\mum}(q)+f_{\mum}(q)(\qm-q)+o(\qm-q)$. Furthermore, notice that $F_{\mum}(\qm)=F_\mu(q)=\alpha$. Then, we have
\[
\qm-q={1\over f_{\mum}(q)}\left( F_\mu(q)-F_{\mum}(q)+o(\qm-q)\right) = {1\over f_{\mum}(q)}\left( \cF_q[\mu]-\cF_q[\mum]+o(\qm-q)\right).
\]
Notice that $f_{\mum}(q)\to f_\mu(q)$ as $m\to\infty$ by Lemma \ref{lem:f} and $\cF_q[\mu]-\cF_q[\mum]=\cO_{\oP}(a_m)$ by Lemma \ref{lem:F}. Then, $\qm-q= \cO_{\oP}(a_m)$.
\hfill$\square$\\[-10pt]     
\end{proof}


\begin{proposition}\label{prop:tm}
Under Assumption \ref{asmp:err_rate} and the regularity conditions of Lemmas \ref{lem:f} to \ref{lem:G}, $\tm-\theta= \cO_{\oP}(a_m)$.
\end{proposition}

\begin{proof}{Proof.}
By the mean value theorem, it can be shown that (see Lemma~\ref{lem:bi-taylor} in  Appendix~\ref{sec:lemma_proof}), as $m\to \infty$, 
\begin{align*}
     G_{\mum,\pim}(\qm,\tm)-G_{\mum,\pim}(q,\theta)&={\partial\over\partial x}G_{\mum,\pim}(q^*,\theta)(\qm-q)+{\partial\over\partial y}G_{\mum,\pim}(q,\theta)(\tm-\theta)+o(\tm-\theta),
\end{align*}
where $q^*$ is some number between $\qm$ and $q$, and hence $q^*\to q$ as $m\to \infty$ due to Proposition~\ref{prop:qm}.
Furthermore, notice that $G_{\mum,\pim}(\qm,\tm)=G_{\mu,\pi}(q,\theta)=\beta$. Then, we have
\begin{eqnarray*}
    \tm-\theta &=& \left[{\partial\over\partial y}G_{\mum,\pim}(q,\theta)\right]^{-1}\cdot \left\{ \left(G_{\mu,\pi}(q,\theta)-G_{\mum,\pim}(q,\theta)\right)- {\partial\over\partial x}G_{\mum,\pim}(q^*,\theta)(\qm-q)+o(\tm-\theta)\right\}\\
    &=& \left[{\partial\over\partial y}G_{\mum,\pim}(q,\theta)\right]^{-1}\cdot \left\{ \left(\cG_{q,\theta}[\mu,\pi]-\cG_{q,\theta}[\mum,\pim]\right)- {\partial\over\partial x}G_{\mum,\pim}(q^*,\theta)(\qm-q)+o(\tm-\theta)\right\},
\end{eqnarray*}
Notice that ${\partial\over\partial x}G_{\mum,\pim}(q^*,\theta)\to {\partial\over\partial x}G_{\mu,\pi}(q,\theta)$ and ${\partial\over\partial y}G_{\mum,\pim}(q,\theta)\to {\partial\over\partial y}G_{\mu,\pi}(q,\theta)$ as $m\to\infty$ by Lemma \ref{lem:g}, $\cG_{q,\theta}[\mu,\pi]-\cG_{q,\theta}[\mum,\pim]=\cO_{\oP}(a_m)$  by Lemma \ref{lem:G}, and $\qm-q= \cO_{\oP}(a_m)$ by Proposition \ref{prop:qm}. Then, $\tm-\theta= \cO_{\oP}(a_m)$.
\hfill$\square$\\[-10pt]    
\end{proof}


Propositions \ref{prop:qm} and \ref{prop:tm} show that the FA error of the CoVaR estimator, i.e., $\tm-\theta$, converges in the same rate as the function-approximation errors of the surfaces, i.e., $\mum-\mu$ and $\pim-\pi$. In particular, we show that it is determined by the convergence rate of the $L^\infty$-norms of $\mum-\mu$ and $\pim-\pi$.

Next, we analyze the convergence rate of the BE-related error, i.e., $\tmn-\tm$. We summarize the result in next proposition, which is built upon Theorem 3 of \cite{huang2022monte}.  

\begin{proposition}\label{prop:BE}
Under Assumption \ref{asmp:err_rate} and the regularity conditions of Lemmas \ref{lem:f} to \ref{lem:G}, when $\sqrt{k}/h\to c$ as $n\to\infty$ for some constant $c>0$, $\tmn-\tm= \cO_{\oP}(n^{-1/3})$.
\end{proposition}

\begin{proof}{Proof.}
Fixing $\mum$ and $\pim$, by Theorem 3 of  \cite{huang2022monte}, we know that when $\sqrt{k}/h\rightarrow c$ for some $c>0$ as $n\rightarrow\infty$,
\[
{\sqrt{k} \cdot {\partial\over\partial y}G_{\mum,\pim}(\qm,\tm) \over \sqrt{\beta(1-\beta)}}(\tmn-\tm)=\cO_{\rm P}(1).
\]
By Lemma \ref{lem:g}, $ {\partial\over\partial y}G_{\mum,\pim}(\qm,\tm)\rightarrow {\partial\over\partial y}G_{\mu,\pi}(q,\theta)$ as $m\rightarrow\infty$.
Therefore, by Slutsky's theorem \citep{serfling2009approximation}, $\tmn-\tm=\cO_{\rm P}(k^{-1/2})=\cO_{\oP}(n^{-1/3})$.
\hfill$\square$\\[-10pt]  
\end{proof}

Let $\Gamma$ denote the number of inner-level observations used by Algorithm \ref{alg:CoVaR}. Notice that $\tmn-\theta=(\tm-\theta)+(\tmn-\tm)$. Then, by Proposition \ref{prop:tm} and \ref{prop:BE}, we can directly obtain the following theorem that characterize the convergence rate of the CoVaR estimator $\tmn$.

\begin{theorem}\label{thm:covar}
    Let $m=\Gamma/l$ where $l\ge 1$ is a constant. Furthermore, let $k$ and $h$ be positive integers that satisfy $kh=n$ and $\sqrt{k}/h\to c$ as $n\to\infty$ for some constant $c>0$. Then, under Assumption \ref{asmp:err_rate} and the regularity conditions of Lemmas \ref{lem:f} to \ref{lem:G}, 
    $\tmn-\theta = \cO_{\oP}(\max\{a_m,n^{-1/3}\})$. If $n$ satisfies $n^{-1/3}\le a_m$, then $\tmn-\theta = \cO_{\oP}(a_m)$.
\end{theorem}

There are several important comments that we want to make regarding Theorem \ref{thm:covar}. First, the conclusion of the theorem is a general result. It may be applied to any smoothing technique as long as it satisfies Assumption \ref{asmp:err_rate}. This is in sharp contrast with the current literature on single-portfolio risk measurement, where estimators are developed for specific smoothing techniques and their generalization to other smoothing techniques are often not straight-forward \citep{liu2010stochastic,broadie2015risk,hong2017kernel,wang2022smooth}. 

Second, the theorem demonstrates that as long as the condition $n^{-1/3}\le a_m$ is satisfied, the convergence rate of the CoVaR estimator is determined by the first stage. Here, $n$ represents the number of second-stage outer-level scenarios, and $m$ is of the same order as $\Gamma$, which is the total number of inner-level observations. Given that outer-level scenarios are much easier to generate than inner-level ones, one should always ensure that this condition is met.

Third, the sampling effort is not the only part of the computational effort that is used to compute the CoVaR estimator. The time used to construct $\mum(\cdot)$ and $\pim(\cdot)$ in the first stage of Algorithm \ref{alg:CoVaR} and the time used to evaluate $\mum(Z_i)$ and $\pim(Z_i)$ for all $i=1,\ldots,n$ in the second stage may also be quite significant, especially for certain smoothing techniques. In the numerical experiments of Section \ref{sec:num}, we compare these times for different smoothing techniques and provide practical recommendations.

Fourth, we want to note that the decoupled algorithm may also be used to estimate the VaR of a single-portfolio. With Proposition \ref{prop:qm} and the central limit theorem of the standard VaR estimator of \cite{serfling2009approximation}, we can easily show that the nested VaR estimator has a convergence rate of $\cO_{\oP}(\min\{a_m,n^{-1/2}\})$. This allows us to reproduce the rate-of-convergence results of most (if not all) smoothing-based nested VaR estimators in the literature, including at least the linear regression estimator of \cite{broadie2015risk}, the kernel estimator of \cite{hong2017kernel} and the KRR estimator of \cite{wang2022smooth}.

\section{Smoothing Techniques}\label{sec:smoothing}

Theorem \ref{thm:covar} shows that the convergence rate of the decoupled CoVaR estimator depends critically on the convergence rate of the smoothing techniques through the $L^\infty$ norms, i.e., $\|\mum-\mu\|_\infty$ and $\|\pim-\pi\|_\infty$, which are determined jointly by two factors, the smoothness of the target functions $\mu$ and $\pi$ and the characteristic of the chosen smoothing technique. In this section, we first demonstrate that the portfolio loss functions employed in CoVaR estimations, namely $\mu$ and $\pi$, generally exhibit high-order smoothness. This observation paves the way for the application of advanced smoothing techniques that can leverage this high-order smoothness. We then derive the rate of convergence for several representative smoothing techniques in terms of the $L^\infty$ norms, based on the smoothness of the portfolio value functions. These results can be incorporated into the convergence rate of the CoVaR estimator through Theorem \ref{thm:covar}, offering valuable insights for selecting appropriate smoothing techniques.

\subsection{Smoothness of Portfolio Loss Functions}\label{subsec:smoothness}

The smoothness of portfolio loss functions is crucial to the decoupled approach for two primary reasons. First, the efficiency of smoothing techniques is inherently linked to the smoothness of the underlying functions, particularly in high-dimensional settings. For instance, smoothness can mitigate the curse of dimensionality when employing KRR for smoothing \citep{wang2022smooth}. Second, extending the convergence rate from the $L^2$ norm, often established for smoothing techniques, to the $L^\infty$ norm, which is necessary for the decoupled approach, is significantly dependent on the smoothness of the functions. This point will be further elaborated in Section \ref{subsec:L_infty results}.

A financial portfolio, comprising multiple derivatives, has a loss function defined as a weighted sum of the loss functions of its individual derivatives. Consequently, the smoothness of the portfolio loss function is contingent upon the smoothness of the price functions of the individual derivatives included in the portfolio. In the remainder of this subsection, we demonstrate that derivative prices typically exhibit high-order smoothness, or even infinite-order smoothness. This characteristic implies that portfolio loss functions also possess high-order smoothness or potentially infinite-order smoothness.

First, we examine three examples in which closed-form expressions for the derivative prices are available. These examples enable us to verify the smoothness and differentiability of the price functions, thereby offering insights into the general case where closed-form expressions may not be attainable.

\begin{example} \label{ex:bs}
Consider a European call option with an underlying asset whose price $S_t$ follows
a geometric Brownian motion, given by ${\rm d} S_t = r S_t {\rm d}t + \sigma S_t {\rm d}B_t$. The payoff function of the option at the expiration $T$ is $(S_T-K)^+$, where $K>0$ is the strike. The price of the option has a closed-form expression $p(S_0)$, included in Appendix \ref{ex:BS_GBM} \citep{black1973pricing}. It is clear that $p(S_0)$ is infinitely differentiable with respect to $S_0$, i.e., $p\in C^\infty$. Note that the payoff function is not differentiable at $S_T=K$. This example demonstrates that the non-differentiability does not affect the smoothness of the price function.
\end{example}

\vspace{5pt}

\begin{example} 
Consider a barrier up-and-out call option with an underlying asset whose price $S_t$ follows a geometric Brownian motion. The payoff of the option at the expiration is $(S_T-K)^+ I{\left\{\max_{t\in(0,T)} S_t < B\right\}}$, where $B>0$ is the barrier. The price of the option has a closed-form expression $p(S_0)$, included in Appendix \ref{ex:barrier} \citep{shreve2004stochastic}. It is clear that $p(S_0)$ is infinitely differentiable with respect to $S_0\in(0,B)$, i.e., $p\in C^\infty((0,B))$. Note that the payoff function has an indicator function in it. This example demonstrates that the discontinuity does not affect the smoothness of the price function.
\end{example}

\vspace{5pt}

\begin{example} Consider a European call option with an underlying asset whose price $S_t$ follows the Heston model (e.g., a stochastic volatility model): ${\rm d} S_t = \mu S_t {\rm d}t + \sqrt{v_t} S_t {\rm d}B^1_t$ and ${\rm d} v_t = \kappa(\theta-v_t) {\rm d}t + \sigma\sqrt{v_t} {\rm d}B^2_t$, where the correlation between two Brownian motions is $corr(B^1_t,B^2_t)=\rho t$. The price of the option has a closed-form expression $p(S_0,v_0)$, included in Appendix \ref{ex:BS_Heston} \citep{heston1993closed}. It can be verified that the price function is infinitely differentiable with respect to both $S_0$ and $v_0$, i.e., $p\in C^\infty$. Note that,  if the option is included in the portfolio, both $S_0$ and $v_0$ are risk factors when measuring CoVaR. This example further extends the smoothness result of Example \ref{ex:bs} to options with multiple risk factors and under more complicated price models.
\end{example}

In practice, closed-form expressions for derivative prices are typically unavailable, highlighting the importance of nested estimation in the assessment of risk measures. In the remainder of this subsection, we demonstrate that the three provided examples are not isolated cases. Under fairly general conditions, portfolio loss functions are infinitely differentiable, even when closed-form expressions of these functions are not obtainable. Rather than analyzing the prices of individual derivatives, we adopt an aggregated perspective, focusing on the smoothness of portfolio loss functions. It is important to note, however, that the two approaches are equivalent.

Consider a general portfolio loss function $\mu(z)=\oE[X \mid Z_0=z]$, where $X=\varphi(Z_T)$ is the aggregated discounted random loss of the portfolio at a future time $T$, $\varphi:\Omega\rightarrow\mR$ is the payoff function, and the expectation is taken with respect to the risk-neutral measure. Typically, the risk factors $Z_T$ are calculated from the prices of underlying assets up to time $T$ discretized at $0<t_1<t_2<\cdots<t_n \leq T$. Suppose there are $q$ assets in the portfolio, and $S_t$ represents their price process, then there is a function $\psi_T:\mR^q\times\cdots\times\mR^q\rightarrow \Omega$ such that $Z_T=\psi_T(S_1,S_2,\cdots,S_n)$, where $S_i\equiv S_{t_i}$.
Suppose $S_t$ is modeled as the solution to the stochastic differential equation
\begin{equation*}
    {\rm d}S_t = \xi(t,S_t){\rm d}t +\Sigma(t, S_t){\rm d}B_t, \quad 0\leq t\leq T,
\end{equation*}
where $\xi:[0,T]\times\mR^q\rightarrow\mR$, $\Sigma:[0,T]\times \mR^q\rightarrow \mR^{q\times q}$ are the coefficient functions, $B_t$ is a standard $q$-dimensional Brownian motion, and $S_0=\psi_0^{-1}(z)$. For $0<i<j< n$, let $p_{i,j}(s,s')$ denote the transition density of $S_j$ at $s'$ given that $S_i=s$.  Then, we have the following proposition showing $\mu(z)$ is infinitely differentiable under some regularity conditions, where we let $\phi(s_1)=\oE[X|S_1=s_1]$ be the portfolio loss once $S_1=s_1$ is observed.

\vspace{5pt}

\begin{proposition}\label{prop:smoothness}
Suppose the following conditions hold:
\begin{itemize}
    \item[(a)] $\oE[|X|\mid Z_0=z]<\infty$;
    \item[(b)] $\psi_0^{-1}(\cdot)$ is smooth;
    \item[(c)] $p_{0,1}(\cdot,s_1)\in C^{\infty}(\mR^q)$;
    \item[(d)] for any $k\geq 1$, $\int |\partial^{\bm k}p_{0,1}(s,s_1)|\phi(s_1){\rm d}s_1<\infty$ for all $s$ and $\int \partial^{\bm k}p_{0,1}(s,s_1)\phi(s_1){\rm d}s_1$
    is continuous function of $s$, where $\bm{k}=(k_1, k_2, \ldots,k_q)\in \mN^q$ with $ |{\bm k}|=k_1+k_2+\cdots+k_d = k$, and $\partial^{\bm k}$ represents the mixed derivative with respect to the entries of $s$.
\end{itemize}

Then, $\mu(z)$ is infinitely differentiable with respect to $z$, i.e., $\mu \in C^{\infty}(\Omega)$.
\end{proposition}

\vspace{5pt}
Proposition \ref{prop:smoothness} is proven using the Leibniz integral rule, as detailed in Appendix~\ref{sec:other-proofs}. Condition (a) is a standard assumption to ensure that the loss surface is well-defined. To illustrate the feasibility of the other conditions in the proposition, consider the following scenario. Assume that the underlying asset prices are included among the risk factors, making up the first $q$ entries of the risk vector. Consequently, $\psi_0^{-1}(z)$ simply extracts these entries, ensuring that condition (b) is satisfied. In addition, assume the drift function $\xi(t,s)$ is affine in $s$, and the diffusion matrix $\Sigma(t,s)$ does not depend on $s$. Under these assumptions, for $S_0 = s$, the variable $S_1$ follows a multivariate normal distribution with a mean vector $b(s)$ linear in $s$ and a constant covariance matrix $A$ \citep{platen2010}. The transition density $p_{0,1}(s, s_1)$ is then given by:
\[
p_{0,1}(s, s_1) = \frac{1}{\sqrt{(2\pi)^d \det(A)}} \exp\left(-\frac{1}{2} (s_1 - b(s))^T A^{-1} (s_1 - b(s))\right),
\]
where its derivatives of any order can be expressed as products of polynomial functions and the normal density. This structure satisfies condition (c). Furthermore, condition (d) holds provided the growth of $\phi$ remains slower than exponential. 

In more general contexts, particularly when the transition density is not directly tractable, verifying the smoothness conditions explicitly may be challenging. However, it is important to note that transition densities, typically modeled as normally distributed when applying the Euler approximation to discretize the asset price process $S_t$, are well-behaved and infinitely differentiable. Thus, these densities generally meet the conditions outlined in Proposition \ref{prop:smoothness}, which confirms that portfolio loss functions are infinitely differentiable. For the remainder of this paper, we will assume that portfolio loss functions demonstrate high-order smoothness, i.e., $\mu \in C^\nu(\Omega)$ and $\pi \in C^\nu(\Omega)$, for some large positive integer $k$, if not being infinitely differentiable.

\subsection{Alternative Smoothing Techniques}\label{subsec:L_infty results}

Let $\Omega$ be a convex and compact subset of $\mR^d$, and let $f:\Omega\rightarrow \mR$ denote the portfolio loss function to be approximated, where $f(z)=\oE[X|Z=z]$. From the analysis in Section \ref{subsec:smoothness}, we know that $f$ is typically sufficiently smooth. Assume that $f$ has smoothness of order of $\nu$, i.e., $f\in C^\nu(\Omega)$, meaning that the derivative $D^{\bm k} f$ exists and is continuous for every $z\in\Omega$, and $\|D^{\bm k} f\|_p<\infty$ for any $|{\bm k}|=k\leq \nu$ and $p>0$, where the $k$-th order derivative $D^{{\bm k}} f$ is defined as 
\[
D^{\bm{k}} f = {\partial^{|{\bm k}|} f\over\partial z_1^{k_1}\partial z_2^{k_2}\ldots \partial z_d^{k_d}}, 
\]
with $\bm{k}=(k_1, k_2, \ldots,k_d)\in \mN^d$ and $ |{\bm k}|=k_1+k_2+\cdots+k_d = k$. We also define $\|D^{k} f\|_p$ as the maximum of the $L^p$ norm of all $k$-th order derivatives of $f$ \citep{nirenberg1959elliptic}. Thus, $D^{\bm{\nu}} f$ represents the mixed partial derivatives of $f$ up to the order $\nu$, and the smoothness condition implies that these derivatives are well-behaved (finite) within $\Omega$. 

Suppose that there are $m$ observations, denoted as $\{(z_1,x_1),(z_2,x_2),\ldots,(z_m,x_m)\}$, where $x_j = f(z_j)+e_j$, and $\{e_1,e_2,\ldots,e_m\}$ is a sequence of independent and identically distributed sub-Gaussian random noises. In this subsection, we introduce four different smoothing techniques—linear regression, kernel smoothing, KRR and neural networks—to approximate the unknown function $f$. Let $\tilde f_m$ denote the approximated function. Our focus is on quantifying the convergence rate of $\|\tilde f_m-f\|_\infty$ as a function of the sample size $m$.

To quantify the convergence rate of $\|\tilde f_m-f\|_\infty$ for KRR and neural networks, we first need to establish a connection with $\|\tilde f_m-f\|_2$, for which the convergence rates are well documented in the literature. For this purpose, we introduce the Gagliardo-Nirenberg interpolation inequality \citep{nirenberg1959elliptic}. Although this inequality has a more general form, we simplify it to fit within our framework.

\begin{proposition}[Gagliardo-Nirenberg Interpolation Inequality]\label{prop:GN}
Suppose that $\|\tilde f_m-f\|_2<\infty$ and its derivatives of order $k$ satisfy $\|D^k (\tilde f_m-f)\|_r<\infty$ for some $r\ge 2$. Then, there exists a constant $c>0$ such that 
\begin{equation}\label{eq:GN}
    \| \tilde f_m-f\|_{\infty} \leq c\|D^k (\tilde f_m-f)\|_{r}^\delta\cdot\|\tilde f_m-f\|_{2}^{1-\delta} + c\|\tilde f_m-f\|_{2}
\end{equation}
when $d\le rk$, where $\delta = {rd \over (r-2)d+2rk}\in [0,1]$.
\end{proposition}

The inequality makes it clear that a fast convergence rate of $\|\tilde f_m-f\|_{2}$ alone may not be sufficient to guarantee a fast convergence rate of $\|\tilde f_m-f\|_{\infty}$. The result also critically depends on the derivatives of order $k$. First, for complex smoothing techniques such as KRR and neural networks, we aim to identify $k$ such that $\|D^k (\tilde f_m-f)\|_{r}<\infty$. Note that $\|D^k (\tilde f_m-f)\|_{r}\le \|D^k f\|_{r}+\|D^k \tilde f_m\|_{r}$. Thus, the convergence rate depends on the smoothness of both the target function $f$ and the approximated function $\tilde f_m$. Second, assuming $\|D^k (\tilde f_m-f)\|_{r}<\infty$, the convergence rate is determined by $\|\tilde f_m-f\|_{2}^{1-\delta}$. A larger $k$ implies a smaller $\delta$, thereby leading to a faster convergence rate. As discussed in Section 4.1, portfolio loss functions generally exhibit high-order smoothness, meaning that $\|D^k f\|_{r}<\infty$ for large $k$. Consequently, the smoothing techniques must ensure that $\tilde f_m$ also has high-order smoothness.

In the remainder of this subsection, we introduce four different smoothing techniques—linear regression, kernel smoothing, KRR and neural networks—and quantify the convergence rate of their $\|\tilde f_m-f\|_{\infty}$.

\vspace{5pt}

\textbf{Linear Regression.} 
Consider a set of $s$ continuous basis functions $b_1(z),\ldots,b_s(z)$ in $\Omega$, and let $b(z)=(b_1(z),\ldots,b_s(z))^\top$. Suppose that $f(z)=\omega^{\top} b(z)+M(z)$, where $\omega\in \mR^{s \times 1}$ is a vector of coefficients and $M(z)$ is the residual bias under the best approximation provided by the basis functions, denoted as $f^\dag(z) = \omega^\top b(z)$. Given the data $\{(z_1,x_1),(z_2,x_2),\ldots,(z_m,x_m)\}$, the estimated coefficients are $\tilde{\omega}_m=\left(B(z)^\top B(z)\right)^{-1}B(z)^\top X$, where $B(z)= [b(z_1), \ldots, b(z_m)]^\top$ and $X = (x_1,\ldots,x_m)^\top$ \citep{james2013introduction}. Then, the approximated function is given by $\tilde f_m(z)=\tilde{\omega}_m^\top b(z)$.

Since $\tilde{\omega}_m-\omega$ is an $s$-dimensional vector, we have 
$$
\|\tilde{f}_m-f^\dag\|_\infty\le \|\tilde{\omega}_m-\omega\|_\infty\cdot\|b\|_\infty\le c\|\tilde{\omega}_m-\omega\|_2 \cdot \|b\|_\infty,
$$ 
where $c>0$ is a constant.
Moreover, because the basis functions are continuous in $\Omega$, and $\Omega$ is a compact set, we know that $\|b\|_\infty<\infty$. By \cite{broadie2015risk}, $\|\tilde{\omega}_m-\omega\|_2 = \mathcal{O}_{\rm P}(m^{-1/2})$. Thus, it follows that $\|\tilde{f}_m-f^\dag\|_\infty = \mathcal{O}_{\rm P}(m^{-1/2})$. Therefore, we have
\[
\|\tilde{f}_m-f\|_\infty
\leq \|\tilde{f}_m-f^\dag\|_\infty + \|f^\dag-f\|_\infty
=\mathcal{O}_{\rm P}(m^{-1/2}) + \|M\|_\infty.
\]
Notice that $m=\Gamma/l$ and $l\ge 1$ is a constant. Consequently, if $n$ is sufficiently large, by Theorem \ref{thm:covar},  $\tmn^{\rm\ LR}$, the CoVaR estimator using linear regression, satisfies 
\[
\tmn^{\rm\ LR}-\theta=\mathcal{O}_{\rm P}\left(\Gamma^{-1/2} + \|M\|_\infty\right).
\]
Therefore, unless the basis functions are chosen such that the residual bias $M(z)=0$ for all $z\in\Omega$, there exists an irreducible error that cannot be eliminated by increasing the sampling effort $\Gamma$. This is consistent with the results of \cite{broadie2015risk} for estimating VaR. In practice, however, if the basis functions are well chosen so that the bias term is small, the linear-regression estimator can exhibit excellent performance.

\vspace{5pt}

\textbf{Kernel Smoothing.} 
Given the data $\{(z_1,x_1),(z_2,x_2),\ldots,(z_m,x_m)\}$, kernel smoothing estimates the function value at $z\in\Omega$ by averaging the $x_j$'s corresponding to the $z_j$'s near $z$, weighted by a kernel-based distance metric. The well-known Nadaraya-Watson kernel estimator, originating from \cite{nadaraya1964estimating} and \cite{watson1964smooth}, is expressed as 
$$
\tilde{f}_m(z)=\frac{\sum_{j=1}^m k_h\left(z-z_i\right)x_j}{\sum_{j=1}^m k_h\left(z-z_j\right)},
$$
where $k_h(t)=\left(1 / h^d\right) k(t / h)$, $k$ is a kernel function, typically a symmetric density function, and $h>0$ is a bandwidth parameter satisfying $h \rightarrow 0$ and $m h^d \rightarrow \infty$ as $m \rightarrow \infty$. For instance, when Gaussian kernel is employed, $k_h(t)={1\over \sqrt{2\pi}h^d}e^{-{t^2\over 2h^2}}$. 

We decompose the estimation error in kernel smoothing into the two parts:
\[
\|\tilde{f}_m-f\|_\infty 
\leq \|\tilde{f}_m-\oE[\tilde{f}_m]\|_\infty + \|\oE[\tilde{f}_m]-f\|_\infty.
\]
By Lemma 2.1 of \cite{hong2017kernel}, $\oE[\tilde{f}_m(z)]-f=\mathcal{O}(h^2)$ for every $z\in\Omega$. Combining this with Theorem 2 from \cite{hansen2008uniform}, which states that $\sup_{z\in\Omega}|\tilde{f}_m-\oE[\tilde{f}_m]|=\mathcal{O}_{\rm P}({ (mh^d)^{-1/2} \sqrt{\log m}})$, we have 
\[
\|\tilde{f}_m-f\|_{\infty}=\mathcal{O}_{\rm P}\left(h^2+(mh^d)^{-1/2} \sqrt{\log m}\right).
\] 
The best rate of convergence is $\mathcal{O}_{\rm P}(({\log m})^{{2\over 4+d}}\,m^{-{2\over 4+d}})$, achieved by selecting the bandwidth $h=\mathcal{O}(({\log m})^{{1\over 4+d}}\,m^{-{1\over 4+d}})$. Consequently, if $n$ is sufficiently large, by Theorem \ref{thm:covar}, $\tilde{\theta}_{m,n}^{\rm\ KS}$, the CoVaR estimator using kernel smoothing, satisfies
\[
\tmn^{\rm\ KS}-\theta=\mathcal{O}_{\rm P}\left(({\log \Gamma})^{{2\over 4+d}}\,\Gamma^{-{2\over 4+d}}\right)\approx \mathcal{O}_{\rm P}\left(\Gamma^{-{2\over 4+d}}\right),
\]
where the approximation holds because we may ignore the logarithmic term as it has little impact on the overall convergence rate.

Kernel smoothing offers a flexible, non-parametric approach for estimating $f$, yet it has limitations. First, it suffers from the curse of dimensionality, as highlighted by \cite{hong2017kernel}, making it inefficient for high-dimensional problems like CoVaR estimation. Second, kernel smoothing is a memory-based technique, requiring the storage and computation of numerous kernel functions to calculate $\tilde f_m(z)$ for each $z$ value. In CoVaR estimation, $\mum(z)$ and $\pim(z)$ need to be computed for many values of $z$ in Stage 2 of Algorithm \ref{alg:CoVaR}, which can be computationally burdensome.

\vspace{5pt}

\textbf{Kernel Ridge Regression}.
KRR, also known as  regularized least-squares \citep{caponnetto2007optimal}, estimates $f$ by solving the following regularized least-squares problem:
$$
\tilde{f}_m=\underset{f \in \mathcal{H}^k}{\arg \min }\ \frac{1}{m} \sum_{j=1}^m \left(f\left(z_j\right)-x_j\right)^2+\lambda\|f\|_{\mathcal{H}^k}^2,
$$
where $\mathcal{H}^k$ denotes the reproducing kernel Hilbert space induced by the kernel function $k(\cdot,\cdot)$, and $\lambda>0$ is a regularization parameter. The solution to this optimization problem has the following representation \citep{scholkopf2001generalized}:
\begin{eqnarray}\label{eq:KRR}
\tilde{f}_m(z) = k_z^\top(K+m\lambda I)^{-1}X,	
\end{eqnarray}
where $k_z=(k(z_1,z),\ldots,k(z_m,z))^\top$, $K$ is an $m\times m$ matrix with $(i,j)$-th entry $k(z_i,z_j)$, and $X=(x_1, \ldots, x_m)^\top$. When using the Mat\'ern kernel with smoothness parameter $\vartheta$, we have $\|\tilde{f}_m\|_2<\infty$ and $\|D^k \tilde{f}_m\|_2<\infty$ for any $0<k\leq \tilde \vartheta=\vartheta+d/2$ \citep{kanagawa2018gaussian}. 

Given that, at the beginning of this subsection, we assumed $f$ has smoothness of order $\nu$, implying that $\|D^k f\|_2<\infty$ for any $k\le \nu$, we can choose $\vartheta$ such that $\tilde \vartheta\ge\nu$. This guarantees that $\|D^\nu(\tilde f_m - f)\|_2\le \|D^\nu \tilde f_m\|_2+\|D^\nu  f\|_2<\infty$. By Proposition \ref{prop:GN}, it is clear that $\delta={d\over2\nu}$, and thus: 
\[
\|\tilde{f}_m-f\|_\infty 
\leq c \|D^\nu(\tilde{f}_m-f)\|_2^\delta\cdot \|\tilde{f}_m-f\|_2^{1-\delta} + c\|\tilde{f}_m-f\|_2
=\mathcal{O}\left(\|\tilde{f}_m-f\|_2^{{2\nu-d \over 2\nu}}\right)
=\mathcal{O}_{\rm P}\left(m^{-{2\nu-d \over 4\nu+2d}}\right),
\]
where the last equality holds because $\|\tilde{f}_m-f\|_2=\mathcal{O}_{\rm P}\left(m^{-{\nu \over 2\nu+d}}\right)$ for KRR, as shown by \cite{steinwart2009optimal}. Consequently, if $n$ is sufficiently large, by Theorem \ref{thm:covar}, $\tilde{\theta}_{m,n}^{\rm\ KRR}$, the CoVaR estimator using KRR, satisfies
\[
\tmn^{\rm\ KRR}-\theta=\mathcal{O}_{\rm P}\left(\Gamma^{-{2\nu-d \over 4\nu+2d}}\right)\approx \mathcal{O}_{\rm P}\left(\Gamma^{-1/2}\right),
\]
where the approximation holds because portfolio loss functions typically exhibit high-order smoothness (i.e., $\nu$ tends to be large, potentially even infinite), as we discussed in Section \ref{subsec:smoothness}. In such cases, the CoVaR estimator overcomes the curse of dimensionality and achieves approximately the best possible rate for Algorithm \ref{alg:CoVaR}. 

However, despite its favorable sample efficiency, KRR is not without limitations. Like kernel smoothing, it is a memory-based technique, requiring the computation, storage, and manipulation of kernel matrices to compute $\tilde{f}_m(z)$ for each $z$, which makes the Stage 2 of Algorithm \ref{alg:CoVaR} computationally expensive. Additionally, fitting $\tilde f_m$ requires tuning hyper-parameters, such as the regularization parameter $\lambda$ and the kernel length scale, a process that is both time-consuming and critical for the algorithm's performance.

\vspace{5pt}

\textbf{Neural Networks.} 
Neural networks are universal function approximators, as established by \cite{hornik1989multilayer}. They learn to map inputs to outputs by processing data through layers of interconnected nodes, analogous to the interactions between neurons in the brain. when using neural networks, the first step is to choose their architecture, which includes selecting the activation function $\sigma(\cdot)$, the number of layers $L$, the number of hidden units $H^l$ for each layer $l=1,\ldots,L$, and the structure of connections between layers, which defines the function class. In this paper, We focus on the popular class of neural networks, multilayer perceptron (MLP) with the sigmoid activation function $\sigma(x)=1/(1+e^{-x})$:
\[
\mathcal{F}(L,h,w) = \{\tilde{f}_{\rm MLP}:  H^1=\cdots=H^L=h,  |w_{ij}^l|\leq w \text{ for all } i, j,l\},
\]
where the MLP is represented as $$\tilde{f}_{\rm MLP} = \bm{W}^L\sigma(\cdots\sigma(\bm{W}^1\sigma(\bm{W}^0 z+\bm{b}^0)+\bm{b}^1)+\cdots)+\bm{b}^L$$
with weight matrices $\bm{W}^l\in\mR^{H^{l+1}\times H^{l}}$  and the bias vector $ \bm{b}^l\in\mR^{H^{l+1}\times 1}$, where $w_{ij}^l$ denotes the elements of weight matrix $\bm{W}^l$.

Given the data $\{(z_1,x_1),(z_2,x_2),\ldots,(z_m,x_m)\}$, neural networks estimate $f$ by solving the following optimization problem:
$$
\tilde{f}_{m} = \underset{f_\Xi \in \mathcal{F}(L,h,w)}
{\arg \min}{1\over m} \sum_{j=1}^m \left(f_\Xi(z_j)-x_j\right)^2,
$$
where $\Xi$ represents the collection of all weights and biases across the layers. Due to the smoothness of the sigmoid activation function and the constraint $|w_{ij}^l|\leq w$ for all $i,j,l$, we have $\|D^k \tilde{f}_{m}\|_\infty<\infty$ for any $k<+\infty$, following the chain rule. Recalling that $f$ is assumed to have smoothness of order $\nu$, i.e., $\|D^k f\|_\infty<\infty$ for any $k\leq\nu$, it follows that $\|D^\nu(\tilde{f}_m-f)\|_\infty \leq \|D^\nu \tilde{f}_{m}\|_\infty + \|D^\nu f\|_\infty<\infty$. By Proposition \ref{prop:GN}, we have $\delta=d/(d+2\nu)$ and 
\begin{eqnarray*}
\|\tilde{f}_{m}-f\|_\infty
&\leq & c\|D^\nu(\tilde{f}_{m}-f)\|_\infty^\delta\cdot\|\tilde{f}_{m}-f\|_2^{1-\delta} + c\|\tilde{f}_{m}-f\|_2\\
&=& \mathcal{O}\left(\|\tilde{f}_{m}-f\|_2^{{2\nu\over2\nu+d}}\right)
=\mathcal{O}_{\rm P}\left((\log m)^{{3\nu\over2(d+2\nu)}}m^{-{2 \nu^2\over (2\nu+d)^2}}\right)
\end{eqnarray*}
where the last equality holds because $\|\tilde{f}_{m}-f\|_2 = \mathcal{O}_{\rm P}\left((\log m)^{3/2} m^{-{\nu\over 2\nu+d }}\right)$, as demonstrated by \cite{langer2021analysis}.  Consequently, if $n$ is sufficiently large, by Theorem \ref{thm:covar}, $\tilde{\theta}_{m,n}^{\rm\ NN}$, the CoVaR estimator using neural networks, satisfies
\[
\tmn^{\rm\ NN}-\theta=\mathcal{O}_{\rm P}\left((\log \Gamma)^{{3\nu\over2(d+2\nu)}}\Gamma^{-{2 \nu^2\over (2\nu+d)^2}}\right)\approx \mathcal{O}_{\rm P}\left(\Gamma^{-1/2}\right),
\]
where the approximation holds because portfolio loss functions typically exhibit high-order smoothness and the logarithmic term is ignored. Similarly to using KRR, the CoVaR estimator using neural networks also overcomes the curse of dimensionality and achieves approximately the best possible rate for Algorithm \ref{alg:CoVaR}. 

Compared to KRR, neural networks require even more time for tuning and training, as it needs to tune more hyper-parameters than KRR. However, they also offer a key advantage over KRR: once trained, they can compute $\tilde f_m(z)$ for any $z$ very quickly, regardless of the size of the training data, as they are not memory-based. This speed makes neural networks ideal for the second stage of Algorithm \ref{alg:CoVaR}, allowing the algorithm to effectively adapt to an offline-online setting, as we will elaborate in Remark \ref{remark:osoa} at the end of this section.

\vspace{6pt}

\textbf{Summary.} Among the four smoothing techniques considered in this section, kernel smoothing suffers from the curse of dimensionality, making it less suitable for CoVaR estimation, which requires approximating high-dimensional portfolio loss functions.  Linear regression is easy to implement and computationally efficient, but it can suffer from bias, particularly in high-dimensional settings where selecting appropriate basis functions may be challenging. Both KRR and neural networks can exploit the high-order smoothness of portfolio loss functions, thereby mitigating the curse of dimensionality. However, both methods are computationally intensive to fit and require extensive parameter tuning. Section \ref{sec:num} presents extensive numerical studies to compare the performance of various smoothing techniques.

It is important to emphasize that Algorithm \ref{alg:CoVaR} and Theorem \ref{thm:covar} provide a flexible ``plug-and-play" framework for estimating CoVaR. This framework allows for the use of any smoothing technique to approximate the portfolio loss functions  $\mu(z)$ and $\pi(z)$, and it offers the convergence rate of the resulting CoVaR estimator, provided the $\|\tilde{f}_m-f\|_\infty$ convergence rate of the smoothing technique is known. Therefore, while we focus on four commonly used smoothing techniques in this section, any smoothing technique with a known convergence rate can be utilized within this framework.

We conclude this section with a remark on the potential of shifting the first stage of Algorithm \ref{alg:CoVaR} to an offline setting, making the heavy computational effort required for generating inner-level observations and training and tuning functional approximations more manageable.

\begin{remark} \label{remark:osoa}
    The decoupled approach offers another interesting potential. In the first stage of the algorithm, full knowledge of the distribution of $Z$ is not necessary, since the goal is simply to approximate the loss functions $\mu(z)$ and $\pi(z)$. This means the function approximations can be done prior to time $0$, before the distribution of $Z$ is known, as long as the portfolio remains unchanged. By constructing $\mum(z)$ and $\pim(z)$ offline, we can employ a larger number of inner-level observations and have more time for training and tuning, both of which will improve the accuracy of the approximations. In the online phase, only the second stage of Algorithm \ref{alg:CoVaR} needs to be implemented, which only involves cheap outer-level scenarios.

    Among the smoothing techniques discussed in Section \ref{subsec:L_infty results}, kernel smoothing and KRR are not ideal for the online phase because they are memory-based methods, which are slow to evaluate $\mum(z)$ and $\pim(z)$ for newly simulated $z$ values. However, linear regression and neural networks are highly efficient for real-time evaluation, making them suitable candidates for the online phase. This efficiency allows for a larger second-stage sample size $n$, leading to better-performing CoVaR estimators.
\end{remark}

\section{Numerical Studies}\label{sec:num}

In this section, we evaluate the empirical performance of the decoupled estimators of CoVaR. Specifically, we focus on three key questions: 
(i) Do decoupled estimators outperform coupled estimators? 
(ii) What are the empirical sample efficiencies of decoupled estimators with different smoothing techniques?
(iii) What are the practical guidelines for selecting an appropriate smoothing technique?
To ensure the practical relevance of the example, we consider two high-dimensional portfolios, each characterized by the same 300-dimensional risk factors, and estimate the corresponding CoVaR. All experiments are implemented in Python and run on a system equipped with two Intel Xeon Gold 6248R CPUs (each with 24 cores) and 256GB of RAM.

We firstly introduce the settings of the example. Let $X$ and $Y$ denote the discounted random losses of the two portfolios at a future time $T$, both underlying the same risk factors. Each portfolio consists of stocks, up-and-out barrier call options, and geometric Asian call options, whose values at time $t$ are written on $q=100$ stocks $\bm{S}(t)=(S_1(t),\ldots,S_q(t))\in\mR^q$. We assume the stock prices follow a multi-dimensional geometric Brownian motion, where for each dimension:
\begin{eqnarray}\label{GBM}
	{dS_i(t)\over S_i(t)}=\mu_i dt+\bar{\sigma}_i d\bar{B}_i(t), \forall i = 1,\ldots,q,
\end{eqnarray}
with $\mu_i$ representing the drift, taking real-world annual return of the stock for $t\in[0,\tau]$ and the risk-free rate $r_f$ for $t\in[\tau,T]$, and $\bar{B}_i(t)$ representing a standard one-dimensional Brownian motion. The Brownian motions $\bar{B}_i(t)$ and $\bar{B}_j(t)$ are correlated with correlation $\rho_{ij}$. Let $\Sigma\in\mR^{q\times q}$ denotes the matrix whose element $\Sigma_{ij}=\rho_{ij}\bar{\sigma}_i\bar{\sigma}_j$. Then, $(\bar{\sigma}_1B_1(t),\ldots,\bar{\sigma}_qB_q(t))\sim BM(0,\Sigma)$. Let $A$ denote a matrix such that $AA^\top=\Sigma$, which can be obtained via the Cholesky factorization of $\Sigma$ \citep{glasserman2004monte}. In this way, we can rewrite Equation \eqref{GBM} as
\begin{eqnarray*}
{dS_i(t)\over S_i(t)}=\mu_i dt+\sum_{j=1}^{q}\sigma_{ij} dB_j(t), \forall i = 1,\ldots,q,	
\end{eqnarray*}
where $\sigma_{ij}$ are the elements of $A$ and $B(t) = (B_1(t),\ldots,B_q(t))$ is a standard multi-dimensional Brownian motion, BM$(0,I)$, where $I$ is the $q$-dimensional identity matrix.

Let $V_0^1$ and $V_0^2$ denote the portfolio values at at $t=0$, and $V_\tau^1$ and $V_\tau^2$ the portfolio values at $t=\tau$. Notice that $V_0^1$ and $V_0^2$ are observable at time $0$. According to derivative pricing theory \citep{duffie2010dynamic}, the portfolio loss function at $t=\tau$ can be expressed as $\mu(Z)=V_0^1-V_\tau^1$ and $\pi(Z)=V_0^2-V_\tau^2$, where
\begin{eqnarray}
V^j_{\tau} &=& \operatorname{E}\left[ w^j_1 \sum_{i=1}^q S_i(\tau)
+w_2^j  \sum_{i=1}^q e^{-r_f(T-\tau)} \left\{\left(\prod_{l=1}^{M} S_i(t_l)\right)^{1/M}-K_{i1}^j\right\}^+\right.\nonumber\\
&& + \left.w_3^j \sum_{i=1}^q e^{-r_f(T-\tau)}\left.(S_i(T)-K_{i2}^j)^+I\left\{\max_{0\leq t\leq T}S_i(t)\leq B_i^j\right\}\right|Z\right], \quad {j=1,2}, \label{eq:V}
\end{eqnarray}
and the vector of risk factors $Z\in\mR^{3q}$  is defined as
 \[
 Z = \left(S_1(\tau),\ldots,S_q(\tau),\max_{0\leq t\leq \tau}S_1(t),\ldots,\max_{0\leq t\leq \tau}S_q(t), \left(\prod_{l=1}^{M_\tau} S_1(t_l)\right)^{1/M_\tau},\ldots,\left(\prod_{l=1}^{M_\tau} S_q(t_l)\right)^{1/M_\tau}\right).
 \]
Here, $K_{i1}^j$, $K_{i2}^j$ and $B_i$ are the strike prices and the barrier prices. 
The Asian options are monitored at time $0<t_1<\ldots<t_M=T$, where $M$ is the times of monitoring during $t\in[0,T]$. We also assume that the risk horizon $\tau=t_{M_\tau}$, with $M_\tau$ times of monitoring during $t\in[0,\tau]$. 
The positions of the two portfolios are represented as
$\bm{w}^1=(w_1^1, w_2^1, w_3^1)$ and $\bm{w}^2=(w_1^2, w_2^2, w_3^2)$, respectively. 

To estimate the portfolio losses at time $\tau$, a two-level simulation is typically conducted: simulating observations of risk factor $Z$ as the outer-level scenarios and then, based on outer-level scenario, generating the inner-level sample paths of the risk factors from time $\tau$ to $T$ to obtain the observations of $X$ and $Y$ according to Equation \eqref{eq:V}. To obtain a more accurate estimator of barrier option price, we apply the Brownian interpolation method discussed in the Section 6.4 of \cite{glasserman2004monte}. Also, considering the high-dimensional nature of this example, we consider three smoothing techniques: linear regression, KRR and neural networks.

Other details of the numerical settings are summarized as follows. We set $q=100$ and thus the risk factor vector $Z$ has $300$ dimensions. The risk free rate is set to $r_f=0.05$, and the drift vector $\mu$ and covariance matrix $\Sigma$ are generated randomly with a fixed random seed. We set $T=1$, divide the period into $50$ intervals, and set $\tau=2/50$. For simplicity, we set $M=50$ as well. The strike prices $K_{i1}^j$ and $K_{i2}^j$ are all set as $105$, and the barrier prices $B_i^j$ are all set to $120$. The position vectors of the two portfolios are $\bm{w}^1=[2,1,-1]$ and $\bm{w}^2=[2,-1,3]$. Furthermore, in this section, we are interested in estimating the $(0.95,0.95)$-CoVaR, i.e., $\alpha=\beta=0.95$. 

It is important to note that, for this particularly example, the closed-form expressions of $\mu(z)$ and $\pi(z)$ may be derived, so we can accurately calculate the true value of the CoVaR and use it as the benchmark. In particular, we find the $(0.95,0.95)$-CoVaR of this example is 261.66. When implementing the CoVaR estimators considered in this paper, we do not use these closed-form expressions, and estimate $\mu(z)$ and $\pi(z)$ through either nested simulation or smoothing.

\subsection{Empirical Sample Efficiency}

In this subsection, we perform numerical experiments to evaluate the performance of various CoVaR estimators under different budget levels of $\Gamma$. Specifically, we consider four coupled estimators: the standard nested simulation (SNS) estimator, along with three smoothing-based coupled estimators employing linear regression (LR), KRR, and neural networks (NN), respectively. These estimators were referred to as ``na\"ive estimators" in Section \ref{subsec:naive}. Additionally, we assess three decoupled estimators that utilize LR, KRR, and NN, respectively. The experiments are conducted across three budget scenarios: $\Gamma=1\times 10^4$, $1\times 10^5$ and $1\times 10^6$.

The performance of the estimators is evaluated using three metrics: relative bias (r-bias), relative standard deviation (r-SD), and relative root mean-squared error (r-RMSE). These metrics are calculated based on 40 replications of the simulation. Let $m$ denote the outer-level sample size in the two-level simulation, and $l$ represent the inner-level sample size for each outer-level scenario, with $\Gamma=ml$ representing the total budget of inner-level observations. For the batching estimation, we use $k$ and $h$ to denote the number of batches and the batch size, respectively. It is important to note that in the case of coupled estimators, the relationship $m=kh$ must be satisfied, whereas in decoupled estimators, $k$ and $h$ are independent of $m$. Therefore, for the decoupled approach, we set $k=h=500$, i.e., $n=2.5\times 10^5$, due to the lower computational cost of the outer-level simulation. The results of the coupled estimators are reported in Tables \ref{table:co_sns} to \ref{table:co_nn}, while the result of the decoupled estimators are presented in Tables \ref{table:de_lr} to \ref{table:de_nn}. 

\begin{table}[ht]
	\captionsetup{labelfont=bf}
	\centering
\caption{The Performance of SNS}\label{table:co_sns}
\begin{tabular}{cccccccc}
\toprule
$\Gamma$       & $m$            & $l$ & $k$ & $h$ & r-bias   & r-SD     & r-RMSE\\\cmidrule{1-3}\cmidrule(l){4-5}\cmidrule(l){6-8}
$1\times 10^4$ & $1\times 10^3$ & 10 & 50         & 20          & 0.2441 & 0.1356 & 0.2793 \\
$1\times 10^5$ & $1\times 10^4$ & 25 & 80         & 50          & 0.1936 & 0.0682 & 0.2052 \\
$1\times 10^6$ & $5\times 10^4$ & 50 & 200        & 100         & 0.1222 & 0.0315 & 0.1262\\\bottomrule
\end{tabular}

\bigskip

\captionsetup{labelfont=bf}
\centering
\caption{The Performance of Coupled LR Approach}\label{table:co_lr}

\begin{tabular}{cccccccc}
\toprule
$\Gamma$       & $m$            & $l$ & $k$ & $h$ & r-bias   & r-SD     & r-RMSE\\\cmidrule{1-3}\cmidrule(l){4-5}\cmidrule(l){6-8}
$1\times 10^4$ & $1\times 10^3$ & 10 & 50         & 20          & 0.1655 & 0.0806 & 0.1841 \\
$1\times 10^5$ & $1\times 10^4$ & 10 & 100         & 100          & 0.0476 & 0.0384 & 0.0612 \\
$1\times 10^6$ & $5\times 10^4$ & 20 & 250        & 200         & 0.0148 & 0.0169 & 0.0225\\\bottomrule
\end{tabular}

\bigskip

\captionsetup{labelfont=bf}
\centering
\caption{The Performance of Coupled KRR Approach}\label{table:co_krr}

\begin{tabular}{cccccccc}
\toprule
$\Gamma$       & $m$            & $l$ & $k$ & $h$ & r-bias   & r-SD     & r-RMSE\\\cmidrule{1-3}\cmidrule(l){4-5}\cmidrule(l){6-8}
$1\times 10^4$ & $1\times 10^3$ & 10 & 50         & 20          & 0.0229 & 0.0597 & 0.0640 \\
$1\times 10^5$ & $1\times 10^4$ & 10 & 100         & 100          & 0.0214 & 0.0368 & 0.0426 \\
$1\times 10^6$ & $5\times 10^4$ & 20 & 250        & 200         & 0.0145 & 0.0175 & 0.0227\\\bottomrule
\end{tabular}
\bigskip

\captionsetup{labelfont=bf}
\centering
\caption{The Performance of Coupled NN Approach}\label{table:co_nn}

\begin{tabular}{cccccccc}
\toprule
$\Gamma$       & $m$            & $l$ & $k$ & $h$ & r-bias   & r-SD     & r-RMSE\\\cmidrule{1-3}\cmidrule(l){4-5}\cmidrule(l){6-8}
$1\times 10^4$ & $1\times 10^3$ & 10 & 50         & 20          & 0.0589 &0.0600 &0.0840  \\
$1\times 10^5$ & $1\times 10^4$ & 10 & 100         & 100          & 0.0251 & 0.0275 & 0.0373 \\
$1\times 10^6$ & $5\times 10^4$ & 20 & 250        & 200         & 0.0255 & 0.0201 & 0.0325\\\bottomrule
\end{tabular}
\end{table}

\begin{table}[ht]
\captionsetup{labelfont=bf}
\centering
\caption{The Performance of Decoupled LR Approach}\label{table:de_lr}

\begin{tabular}{cccccccc}
\toprule
$\Gamma$       & $m$            & $l$ & $k$ & $h$ & r-bias   & r-SD     & r-RMSE\\\cmidrule{1-3}\cmidrule(l){4-5}\cmidrule(l){6-8}
$1\times 10^4$ & $1\times 10^3$ & 10 & 500        & 500         & 0.2155 & 0.0383 & 0.2189 \\
$1\times 10^5$ & $1\times 10^4$ & 10 & 500        & 500         & 0.0338 & 0.0154 & 0.0371 \\
$1\times 10^6$ & $5\times 10^4$ & 20 & 500        & 500         & 0.0061 & 0.0138 & 0.0151\\\bottomrule
\end{tabular}

\bigskip

\captionsetup{labelfont=bf}
\centering
\caption{The Performance of Decoupled KRR Approach}\label{table:de_krr}

\begin{tabular}{cccccccc}
\toprule
$\Gamma$       & $m$            & $l$ & $k$ & $h$ & r-bias   & r-SD     & r-RMSE\\\cmidrule{1-3}\cmidrule(l){4-5}\cmidrule(l){6-8}
$1\times 10^4$ & $1\times 10^3$ & 10 & 500        & 500         & -0.0151& 0.0270 & 0.0309 \\
$1\times 10^5$ & $1\times 10^4$ & 10 & 500        & 500         & 0.0037 & 0.0132 & 0.0137 \\
$1\times 10^6$ & $5\times 10^4$ & 20 & 500        & 500         & 0.0015 & 0.0103 & 0.0104\\\bottomrule
\end{tabular}

\bigskip

\captionsetup{labelfont=bf}
\centering
\caption{The Performance of Decoupled NN Approach}\label{table:de_nn}

\begin{tabular}{cccccccc}
\toprule
$\Gamma$       & $m$            & $l$ & $k$ & $h$ & r-bias   & r-SD     & r-RMSE\\\cmidrule{1-3}\cmidrule(l){4-5}\cmidrule(l){6-8}
$1\times 10^4$ & $1\times 10^3$ & 10 & 500        & 500         & 0.0168 & 0.0317 & 0.0359 \\
$1\times 10^5$ & $1\times 10^4$ & 10 & 500        & 500         & 0.0021 & 0.0121 & 0.0123 \\
$1\times 10^6$ & $5\times 10^4$ & 20 & 500        & 500         & 0.0088 & 0.0126 & 0.0154\\\bottomrule
\end{tabular}
\end{table}

From these numerical results, we can address the first two questions posed at the beginning of this section. First, the decoupled approach clearly outperforms the coupled approach across the board. The standard nested simulation estimator is inefficient for estimating CoVaR, with a relative RMSE of 12.62\% even when $\Gamma = 10^6$. Regardless of the smoothing technique applied, decoupled estimators consistently achieve better sample performance than coupled estimators. For instance, with $\Gamma = 10^5$ for KRR and neural networks, or $\Gamma = 10^6$ for linear regression, decoupled estimators yield relative RMSEs below 2\%, a level of accuracy that none of the coupled estimators can match. This result underscores the significant advantage of decoupled approach.

Secondly, as the computational budget increases, nearly all performance metrics improve, aligning with our theoretical expectations. However, several interesting observations emerge. While linear regression under-performs KRR and neural networks with a smaller budget ($\Gamma = 1 \times 10^4$), it achieves excellent results with a sufficiently large dataset ($\Gamma = 1 \times 10^6$), obtaining a relative RMSE of just 1.51\%. Even with a small computational budget ($\Gamma = 10^4$), KRR and neural networks deliver strong CoVaR estimates, with relative RMSEs below 3.6\%. However, increasing computational budget does not always yield better results, as seen in the last line of Table \ref{table:de_nn}. The empirical performance of KRR and neural networks is highly sensitive to the selection of hyper-parameters. A larger dataset significantly slows down the hyper-parameter tuning process, making it more difficult to identify optimal settings. 

While inner-level simulation is often regarded as the primary computational bottleneck, smoothing techniques that require extensive tuning and iterative training, such as KRR and neural networks, can also add considerable computational overhead. Therefore, it is crucial to consider these trade-offs when assessing the overall computational efficiency of different estimators. We will explore this issue in more detail in the next subsection.

\subsection{Computational Efficiency and Selection of Smoothing Techniques}

In this subsection, we focus on the computational efficiency of decoupled estimators and aim to provide intuitive guidelines for selecting appropriate smoothing techniques. We break down the entire estimation process of the decoupled approach into more detailed phases and carefully analyze the computation time for each. Specifically, the first stage is split into three phases: two-level simulation, tuning, and fitting. The second stage is divided into two phases: simulation and estimation.

It's worth recalling that J.P. Morgan’s well-known ``4:15 report" requires the company to consolidate the risks from all trading desks within 15 minutes of the market’s closing, based on the final values of the underlying factors. In this numerical example, we strive to keep the computation time for decoupled approaches, using linear regression, KRR, and neural networks, within that 15-minute window. The sample allocation, estimation error, and computation time for each phase are reported in Tables \ref{table:time_lr} to \ref{table:time_nn}, which are based on 40 replications.

\begin{table}[ht]
\captionsetup{labelfont=bf}
\centering
\caption{The Computation Time of Decoupled LR}\label{table:time_lr}
\begin{tabular}{cccccccc}
\toprule
\multicolumn{5}{c}{Sample Allocation}               & \multicolumn{3}{c}{Error} \\\cmidrule{1-5}\cmidrule(l){6-8}
$\Gamma$       & $m$            & $l$ & $k$  & $h$  & r-bias    & r-SD     & r-RMSE   \\\cmidrule{1-3}\cmidrule(l){4-5}\cmidrule(l){6-8}
$1\times 10^6$ & $5\times 10^4$ & 20  & 1500 & 1500 & -0.0012  & 0.0074 & 0.0075\\\bottomrule
\end{tabular}

\smallskip

\begin{tabular}{cccccc}
\toprule
\multicolumn{6}{c}{Computation Time}\\\midrule
\multicolumn{3}{c}{The First Stage} & \multicolumn{2}{c}{The Second Stage} &        \\\cmidrule{1-3}\cmidrule(l){4-5}\cmidrule(l){6-6}
simulation   & tuning   & fitting   & simulation        & estimation       & total  \\\cmidrule{1-3}\cmidrule(l){4-5}\cmidrule(l){6-6}
615.20      & 193.04   & 1.61      & 70.03             & 29.20            & 909.08\\\bottomrule
\end{tabular}

\bigskip

\captionsetup{labelfont=bf}
\centering
\caption{The Computation Time of Decoupled KRR}\label{table:time_krr}
\begin{tabular}{cccccccc}
\toprule
\multicolumn{5}{c}{Sample Allocation}               & \multicolumn{3}{c}{Error} \\\cmidrule{1-5}\cmidrule(l){6-8}
$\Gamma$       & $m$            & $l$ & $k$  & $h$  & r-bias    & r-SD     & r-RMSE   \\\cmidrule{1-3}\cmidrule(l){4-5}\cmidrule(l){6-8}
$1.5\times 10^5$ & $1.5\times 10^4$ & 10  & 500 & 500 & 0.0049 & 0.0145 & 0.0153 \\\bottomrule
\end{tabular}

\smallskip

\begin{tabular}{cccccc}
\toprule
\multicolumn{6}{c}{Computation Time}\\\midrule
\multicolumn{3}{c}{The First Stage} & \multicolumn{2}{c}{The Second Stage} &        \\\cmidrule{1-3}\cmidrule(l){4-5}\cmidrule(l){6-6}
simulation   & tuning   & fitting   & simulation        & estimation       & total  \\\cmidrule{1-3}\cmidrule(l){4-5}\cmidrule(l){6-6}
91.73 & 588.86 & 30.98 & 7.69 & 190.61 & 909.88 \\\bottomrule
\end{tabular}

\bigskip

\captionsetup{labelfont=bf}
\centering
\caption{The Computation Time of Decoupled NN}\label{table:time_nn}
\begin{tabular}{cccccccc}
\toprule
\multicolumn{5}{c}{Sample Allocation}               & \multicolumn{3}{c}{Error} \\\cmidrule{1-5}\cmidrule(l){6-8}
$\Gamma$       & $m$            & $l$ & $k$  & $h$  & r-bias    & r-SD     & r-RMSE   \\\cmidrule{1-3}\cmidrule(l){4-5}\cmidrule(l){6-8}
$5\times 10^4$ & $5\times 10^3$ & 10  & 2000 & 2000 & 0.0011 & 0.0158 & 0.0158 \\\bottomrule
\end{tabular}

\smallskip

\begin{tabular}{cccccc}
\toprule
\multicolumn{6}{c}{Computation Time}\\\midrule
\multicolumn{3}{c}{The First Stage} & \multicolumn{2}{c}{The Second Stage} &        \\\cmidrule{1-3}\cmidrule(l){4-5}\cmidrule(l){6-6}
simulation   & tuning   & fitting   & simulation        & estimation       & total  \\\cmidrule{1-3}\cmidrule(l){4-5}\cmidrule(l){6-6}
28.72 & 615.45 & 130.82 & 114.99 & 22.33 & 912.31   \\\bottomrule
\end{tabular}
\end{table}

To fully utilize the 15-minute time budget, the sample and computation time allocations vary depending on the smoothing technique used. It's essential to note that while inner-level simulations have traditionally been viewed as highly time-consuming, advances in computational power—particularly through matrix operations—have significantly accelerated these processes. In this study, the shift from loop-based calculations (such as in \cite{wang2022smooth}) to matrix operations has led to approximately a fifteenfold increase in the speed of inner-level simulations.

For linear regression, the tuning process involves selecting basis functions via cross-validation. Linear regression tends to perform best with larger sample sizes, and given the simplicity of tuning and fitting, we allocate most of the computation time to the first stage simulation, setting $\Gamma=10^6$. With fast outer-level simulations and rapid prediction from the fitted model, we use $k=h=1500$. This leads to a low relative RMSE of $0.75\%$, as shown in Table \ref{table:time_lr}.

For KRR, tuning involves selecting the penalty coefficient $\lambda$ and the parameters of the kernel function. We use a Gaussian kernel and apply grid search during tuning. While KRR performs well with moderate sample sizes ($\Gamma$), tuning, fitting, and prediction require handling large matrices, which becomes time-intensive and memory-consuming, especially with large $\Gamma$ and $n=kh$. Therefore, we choose $\Gamma=1.5\times10^5$ and $k=h=500$, smaller than for linear regression. As shown in Table \ref{table:time_krr}, most computation time is dedicated to the tuning and estimation phases, resulting in a $1.53\%$ relative RMSE.

For neural networks, tuning is even more complex, requiring the selection of numerous hyper-parameters-such as network architecture, learning rate, dropout rate, and training batch size—using iterative algorithms. Excessively large datasets can hinder performance in limited time windows. Thus, we set $\Gamma=5\times 10^4$ to manage the tuning and training effectively. However, prediction is very fast, allowing us to set a larger sample size for the second stage ($k=h=2000$). Estimation with neural networks is observed to be even faster than with linear regression. The results are presented in Table \ref{table:time_nn}.

Each smoothing technique exhibits different trade-offs between computational time and sample efficiency. For applications where sample generation is not a concern, linear regression is highly efficient, particularly if structural information about the portfolios is available. However, in scenarios with limited computational capacity for simulations, KRR and neural networks can provide more accurate estimations. In cases where the first stage can be performed offline, linear regression and neural networks stand out due to their fast second-stage estimation. 

\section{Conclusions} \label{sec:conclusions}

This paper investigates the estimation of CoVaR in scenarios where the loss functions of the financial portfolios must be estimated from simulation samples. Our approach addresses two primary challenges in this problem: ZPE and repricing. We propose a decoupled approach that incorporates the advanced smoothing techniques, and develop a model-independent theoretical framework for examining the convergence behavior of the proposed estimator. The analysis reveals that the decoupled estimator attains a convergence rate of approximately $\cO_{\rm P}(\Gamma^{-1/2})$ when the computational budget is $\Gamma$. This good sample efficiency is achievable provided the loss functions are sufficiently smooth and the chosen smoothing technique exhibits high $L^\infty$ convergence rates. We further discuss the smoothness of portfolio loss functions under general conditions and the $L^\infty$ convergence rates of various smoothing techniques. Through extensive numerical experiments, we demonstrate the performance of the decoupled estimators of CoVaR, and provide practical insights on selecting appropriate smoothing techniques.



%

\begin{APPENDICES}
\section{Examples of Derivatives Prices With Closed-form Expressions}

\subsection{Example 1: European Call Option Under Geometric Brownian Motion}\label{ex:BS_GBM}
The analytical pricing formula of  European call option with an underlying asset whose price $S_t$ follows
a geometric Brownian motion, given by
\[
 {\rm d} S_t = r S_t {\rm d}t + \sigma S_t {\rm d}B_t
 \]
is given by
\[
C(S_t,t) = S_tN(d_1)-Ke^{-r(T-t)}N(d_2),
\]
where 
\begin{eqnarray*}
d_1 &=& \frac{{\rm ln}(S_t/K)+(r+\sigma^2/2)(T-t)}{\sigma\sqrt{T-t}}\\
d_2 &=& d_1-\sigma\sqrt{T-t},
\end{eqnarray*}
and  $N(\cdot)$ is the standard normal cumulative distribution function.

\subsection{Example 2: Barrier Up-and-Out Call Option Under Geometric Brownian Motion}\label{ex:barrier}
The closed-form pricing expression of barrier up-and-out call option, with an underlying asset whose price $S_t$ follows a geometric Brownian motion, is given by
\begin{equation*}
\begin{aligned}
& C(S_t,t)= S_t\left[N\left(\delta_{+}\left(T-t, \frac{S_t}{K}\right)\right)-N\left(\delta_{+}\left(T-t, \frac{S_t}{B}\right)\right)\right] \\
&-e^{-r (T-t)} K\left[N\left(\delta_{-}\left(T-t, \frac{S_t}{K}\right)\right)-N\left(\delta_{-}\left(T-t, \frac{S_t}{B}\right)\right)\right] \\
&-B\left(\frac{S_t}{B}\right)^{-\frac{2 r}{\sigma^2}}\left[N\left(\delta_{+}\left(T-t, \frac{B^2}{K S_t}\right)\right)-N\left(\delta_{+}\left(T-t, \frac{B}{S_t}\right)\right)\right] \\
&+e^{-r (T-t)} K\left(\frac{S_t}{B}\right)^{-\frac{2 r}{\sigma^2}+1}\left[N\left(\delta_{-}\left(T-t, \frac{B^2}{K S_t}\right)\right)-N\left(\delta_{-}\left(T-t, \frac{B}{S_t}\right)\right)\right], \\
& 0 \leq t<T, 0<S_t < B,
\end{aligned}
\end{equation*}
where
\[
\delta_\pm(\tau,s)={1\over\sigma\sqrt{\tau}}\left[\ln s+(r\pm {1\over 2}\sigma^2)\tau\right].
\]
\subsection{Example 3: European Call Option Under Heston Model}\label{ex:BS_Heston}
Consider a European call option with an underlying asset whose price $S_t$ follows the Heston model as follows,	
\begin{eqnarray*}
{\rm d} S_t &=& \mu S_t {\rm d}t + \sqrt{v_t} S_t {\rm d}B^1_t\\
{\rm d} v_t &=& \kappa(\theta-v_t) {\rm d}t + \sigma\sqrt{v_t} {\rm d}B^2_t,	
\end{eqnarray*}
where the correlation $corr(B^1_t,B^2_t)=\rho t$.

Then, the closed-form expression of the price of European call option is 
\[
C(S_t,v_t,t)=S_tP_1-K e^{r(T-t)}P_2,
\]
where 
\[
P_j(S_t, v_t, T ; \ln K)=\frac{1}{2}+\frac{1}{\pi} \int_0^{\infty} \operatorname{Re}\left[\frac{e^{-i \phi \ln K} f_j(\ln S_t, v_t, T ; \phi)}{i \phi}\right] d \phi, \quad j=1,2.
\]
Specifically, 
$$
f_j(\ln S_t, v_t, t ; \phi)=e^{C(T-t ; \phi)+D(T-t ; \phi) v_t+i \phi \ln S_t},
$$
where
$$
\begin{aligned}
& C(\tau ; \phi)=r \phi i \tau+\frac{a}{\sigma^2}\left\{\left(b_j-\rho \sigma \phi i+d\right) \tau-2 \ln \left[\frac{1-g e^{d \tau}}{1-g}\right]\right\}, \\
& D(\tau ; \phi)=\frac{b_j-\rho \sigma \phi i+d}{\sigma^2}\left[\frac{1-e^{d \tau}}{1-g e^{d \tau}}\right]
\end{aligned}
$$
and
$$
\begin{aligned}
g & =\frac{b_j-\rho \sigma \phi i+d}{b_j-\rho \sigma \phi i-d}, \\
d & =\sqrt{\left(\rho \sigma \phi i-b_j\right)^2-\sigma^2\left(2 u_j \phi i-\phi^2\right)} .
\end{aligned}
$$
Other parameters are defined as
\[
u_1 = 1/2, \quad u_2 = -1/2,\quad a = \kappa\theta,\quad b_1 = \kappa+\lambda-\rho\sigma,\quad b_2 = \kappa+\lambda.
\]
Note that the integrand $P_j(S_t, v_t, T ; \ln K)$ is a smooth function \citep{heston1993closed}.

\section{Lemmas and Proofs}\label{sec:lemma_proof}

\subsection{Lemma~\ref{lem:f} and Its Proof}
\begin{customthm}{\ref{lem:f}}
Suppose Assumption~\ref{asmp:err_rate} holds. In addition, for sufficiently large $m$ and conditioned on fitted portfolio loss functions $\mum(\cdot)$,  the joint density function $b_m(x,u)$ of  $(\mu(Z),\eta_m(Z))$ and its partial derivatives ${\partial_x}b_m(x,u)$ and ${\partial_x^2}b_m(x,u)$ exist for all $x$ and $u$, and satisfies the following regularity conditions: there exist non-negative functions $\bar{b}_{m,i}(u), i=0,1,2$ such that
\begin{equation}\label{eq:asm2_1}
b_m(x,u)\leq \bar{b}_{m,0}(u), \quad \left|{\partial_x}b_m(x,u)\right|\leq \bar{b}_{m,1}(u), \quad \left|{\partial_x^2}b_m(x,u)\right|\leq \bar{b}_{m,2}(u)
\end{equation}
for all $x$ and $u$, with
\begin{equation}\label{eq:asm2_2}
B_{r,i}\equiv\sup_m\int_{\mR}|u|^r\bar{b}_{m,i}(u)<\infty
\end{equation}
for $r=0,1,2,3$ and $i=0,1,2$.
Then for any sequence $\{x_m\}$ such that $x_m\rightarrow x$ as $m\to\infty$, we have
 $f_{\mum}(x_m)\rightarrow f_\mu(x)$  and $f'_{\mum}(x_m)\rightarrow f'_{\mu}(x)$ almost surely as $m\to\infty$.
\end{customthm}

\begin{remark}
The regularity conditions specified in Lemma~\ref{lem:f} and subsequent lemmas within this section align with those typically assumed in the nested simulation literature, e.g., \cite{gordy2010nested}. These conditions are expected to hold when the portfolio loss surfaces and their fitted approximations exhibit sufficient smoothness, aligning with the basic assumptions of our framework. Through our analysis, we assume that boundedness conditions are satisfied almost surely in line with the conventional practices in the literature. However, it is possible to modify these conditions to allow for stochastic boundedness, which would change our results from convergence almost surely to convergence in probability. Although this modification is feasible, it would significantly complicate the proofs without providing substantial additional insights or benefits. Therefore, we retain the original assumptions for clarity and conciseness in our analysis.
\end{remark}

\proof{Proof of Lemma~\ref{lem:f}.}
We represent $f_{\mum}$ and $f_{\mu}$ in terms of $b_m$. Notice that
\begin{eqnarray*}
F_{\mum}(x)&=&\operatorname{Pr}_m\{\mum(Z)\leq x\} 
= \operatorname{Pr}_m\{\mu(Z)+a_m\eta_m(Z)\leq x\}
=\int_{\mR}\int_{-\infty}^{x-a_m u}b_m(s,u)dsdu,\\
F_{\mu}(x)&=&\operatorname{Pr}\{\mu(Z)\leq x\}
=\int_{\mR}\int_{-\infty}^{x}b_m(s,u)dsdu.
\end{eqnarray*}
Notice that for any $x'\neq x$, by \eqref{eq:asm2_1},
\begin{align*}
    \left|\frac{\int_{-\infty}^{x'-a_m u}b_m(s,u)ds-\int_{-\infty}^{x-a_m u}b_m(s,u)ds}{x'-x} \right|= \left|\frac{\int_{x-a_m u}^{x'-a_m u}b_m(s,u)ds}{x'-x} \right| \leq \bar b_{m,0}(u),
\end{align*}
where $\bar b_{m,0}$ is integrable due to \eqref{eq:asm2_2}. Therefore, by dominated convergence theorem, we can derive $f_{\mum}(x)$ by differentiating $F_{\mum}(x)$ and switch integration and differentiation, which yields
\begin{align*}
    f_{\mum}(x) &=\frac{d}{dx}\int_{\mR}\int_{-\infty}^{x-a_m u}b_m(s,u)dsdu= \int_{\mR}\frac{d}{dx}\int_{-\infty}^{x-a_m u}b_m(s,u)dsdu= \int_{\mR}b_m(x-a_mu,u)du.
\end{align*}
Similarly, we have
\begin{eqnarray*}
f_{\mu}(x) &=&\int_{\mR}b_m(x,u)du.
\end{eqnarray*}
As a result,
\begin{eqnarray*}
f_{\mum}(x_m)-f_{\mu}(x)
&=&\int_{\mR}[b_m(x_m-a_mu,u)-b_m(x,u)]du\\
&=& \int_{\mR}{\partial_x} b_m(x^*,u)\cdot(x_m-a_mu-x)du\\
&=& (x_m-x)\int_{\mR}{\partial_x} b_m(x^*,u)du-a_m\int_{\mR} u\cdot{\partial_x} b_m(x^*,u)du.
\end{eqnarray*}
where $x^*$ lies between $x-a_mu$ and $x$. Therefore
\begin{align*}
    |f_{\mum}(x_m)-f_{\mu}(x)| &\leq |x_m-x|\cdot\int_{\mR}\left|{\partial_x} b_m(x^*,u)\right|du+a_m\cdot\int_{\mR}\left|u\cdot {\partial_x} b_m(x^*,u)\right|du\\
    &\leq |x_m-x|\cdot\int_{\mR} \bar b_{m,1}(u)du+a_m\cdot\int_{\mR}|u|\bar b_{m,1}(u)du\\
    &\leq |x_m-x|\cdot B_{0,1}+a_m\cdot B_{1,1}.
\end{align*}
As $m\rightarrow0$, $x_m-x\rightarrow0$ and $a_m\rightarrow0$. Therefore, by \eqref{eq:asm2_2}, we have $f_{\mum}(x_m)\rightarrow f_{\mu}(x)$. Similarly, we can show that $f'_{\mum}(x_m)\rightarrow f'_{\mu}(x)$.
\Halmos
\endproof

\subsection{Lemma~\ref{lem:F} and Its Proof}
\begin{alemma}\label{lem:df_cont} 
Let $\Phi_m(x;w)$ and $\phi_m(x;w)$ be the cumulative distribution function and density of $\mu(Z)+w\eta_m(Z)$ for any $w\in[0,a_m]$. Assume Assumption~\ref{asmp:err_rate} and regularity conditions of Lemma~\ref{lem:f} hold. In addition, assume $f_{\mu}(q)> 0$. Then
\begin{itemize}
\item[(1)] $\phi_m(x;w)$ is continuous with respect to $w$ and $x$.
\item[(2)] $\partial_w\Phi_m(x;w)$ exists and is continuous with respect to $w$ and $x$.
\item[(3)] For any $w\in(0,a_m)$, $\oE_m[\eta_m(Z)|\mu(Z)+w\eta_m(Z)=x]$ is continuous with respect to $x$ at $x=q$ for sufficiently large $m$.
\end{itemize}
\end{alemma}
\proof{Proof of Lemma~\ref{lem:df_cont}.}
We first represent $\Phi_m(x;w)$ and $\phi_m(x,w)$ in terms of $b_m$ by
\begin{align*}
\Phi_m(x;w)&= \operatorname{Pr}_m\{\mu(Z)+w\eta_m(Z)\leq x\}
=\int_{\mR}\int_{-\infty}^{x-w u}b_m(s,u)dudv,\\
\phi_m(x;w)&={\partial_x}\Phi_m(x;w)=\int_{\mR}b_m(x-wu,u)du,
\end{align*}
where the formula for $\phi_m$ is derived by switching integration and differentiation under the regularity conditions on $b_m$ and $\partial_x b_m$ in the similarly way as in the proof of Lemma~\ref{lem:f}. Then apply mean value theorem, and use \eqref{eq:asm2_1} and \eqref{eq:asm2_2}, we have
\begin{align*}
    |\phi_m(x;w')-\phi_m(x;w)|&= \left|\int_{\mR}b_m(x-w'u,u)-b_m(x-wu,u)du\right|\\
     &=\left|\int_{\mR}\partial_xb_m(s^{**},u)(w'-w)(-u)du\right|\\
     &\leq|w'-w|\int_{\mR}|u|\bar b_{m,1}(u)du\\
    &\leq|w'-w|\cdot B_{1,1}.\\
    |\phi_m(x';w)-\phi_m(x;w)|&= \left|\int_{\mR}b_m(x'-wu,u)-b_m(x-wu,u)du\right|\\
     &=\left|\int_{\mR}\partial_xb_m(s^{***},u)(x'-x)du\right|\\
     &\leq|x'-x|\int_{\mR}\bar b_{m,1}(u)du\\
      &\leq|x'-x|\cdot B_{0,1},
\end{align*}
where $s^{**}$ lies between $x-wu$ and $x-w'u$ and $s^{***}$ lies between $x-wu$ and $x'-wu$.
This implies that $\phi_m(x;w)$ is continuous with respect to both $w$ and $x$.

Next, notice that
\begin{align*}
    \left|\frac{\int_{-\infty}^{x-w' u}b_m(s,u)ds-\int_{-\infty}^{x-w u}b_m(s,u)ds}{w'-w} \right|= \left|\frac{\int_{x-w' u}^{x-w u}b_m(s,u)ds}{w'-w} \right| \leq |u|\bar b_{m,0}(u),
\end{align*}
Therefore, by dominated convergence theorem and \eqref{eq:asm2_2}, we can derive $\partial_w\Phi_m(x;w)$ by differentiating $\Phi_m(x;w)$ and switch integration and differentiation, which yields
\begin{align*}
    \partial_w\Phi_m(x;w) &=\partial_w\int_{\mR}\int_{-\infty}^{x-w u}b_m(s,u)dudv= \int_{\mR}\partial_w\int_{-\infty}^{x-w u}b_m(s,u)dudv= -\int_{\mR}u\cdot b_m(x-wu,u)du.
\end{align*}
Similarly, apply mean value theorem, and use \eqref{eq:asm2_1} and \eqref{eq:asm2_2}, we know $\partial_w\Phi_m(x;w)$ is continuous with respect to $w$ and $x$ under the regularity conditions on $u^2 \partial_x b_m$ and $u\partial_x b_m$, respectively.

Last, we have
\begin{align*}
    \oE_m[\eta_m(Z)|\mu(Z)+w\eta_m(Z)=x] =\frac{\int_{\mR} u\cdot b_m(x-wu,u)du}{\phi_m(x,w)}.
\end{align*}
We have shown that both the denominator and the numerator are continuous with respect to $x$. In addition, $\phi_m(x,w)$ is continuous with respect to $w$ and $\phi_m(q,0)\equiv f_\mu(q)> 0$, and hence  $\phi_m(x,w)> 0$ for $x=q$ and $w\in(0,a_m)$ for sufficiently large $m$. Therefore the conditional expectation is continuous at with respect to $x$ at $x=q$ for sufficiently large $m$.
\Halmos
\endproof

\begin{customthm}{\ref{lem:F}}
Under Assumption~\ref{asmp:err_rate} and regularity conditions of Lemma~\ref{lem:f} and \ref{lem:df_cont}, we have
\begin{equation*}
    \cF_{q}[\mum]-\cF_{q}[\mu] =\cO_{\oP}(a_m).
\end{equation*}
\end{customthm}
\proof{Proof of Lemma~\ref{lem:F}.}
We first show that $\cF_{q}[\mu+w\eta_m]$ is differentiable with respect to $w\in(0,a_m)$, and
\begin{equation}\label{eq:dF}
\frac{d}{dw}\cF_{q}[\mu+w\eta_m]=-\phi_m(q;w)\operatorname{E}_m[\eta_m(Z)|\mu(Z)+w\eta_m(Z)=q ].
\end{equation}
This result can be proved by applying Theorem 1 of \cite{hong2009estimating}, which states that for real number $q$ and bi-variate function $h:\Omega\times (0,a_m)\rightarrow\mR$, we have
\begin{equation}\label{eq:proof_dF}
p_q^\prime(w)=-\phi(q;w)\operatorname{E}[\partial_w h(Z,w)|h(Z,w)=q ], 
\end{equation}
for any $w\in(0,a_m)$, where $p_q(w)=\operatorname{Pr}\{h(Z,w)\leq q\}$, if the following conditions are satisfied:
\begin{itemize}
\item[(a)] The path-wise derivative $\partial_w h(Z,w)$ exists w.p.1 for any $w\in(0,a_m)$, and there exists a function $k(Z)$ with $\operatorname{E}[k(Z)]<\infty$, such that $|h(Z,w_1)-h(Z,w_2)|\leq k(Z)|w_2-w_1|$ for all $w_1, w_2$.
\item[(b)] Let $\Phi(t;w)$ and $\phi(t;w)$ denote the cumulative distribution function and density of $ h(Z,w)$.
For any $w\in(0,a_m)$, $ h(Z,w)$  has a continuous density $\phi(t;w)$ in a neighborhood of $t = q$, and $\partial_w \Phi(t;w)$ exists and is continuous with respect to both $w$ and $t$ at $t=q$.
\item[(c)] For any $w\in(0,a_m)$, $\operatorname{E}[\partial_w h(Z,w)|h(Z,w)=t ]$ is continuous at $t=q$.
\end{itemize}
In order to prove \eqref{eq:dF}, we set $h(z,w)=\mu(z)+w\eta_m(z)$, and verify the validity of the above conditions. On the one hand, $\partial_w h(Z,w)=\eta_m(Z)$, and 
$|h(Z,w_1)-h(Z,w_2)|=|\eta_m(Z)(w_2-w_1)|$. Therefore condition (a) holds with $k(Z)=\eta_m(Z)$. On the other hand, condition (b) and (c) also hold due to Lemma~\ref{lem:df_cont}. Therefore, we can apply \eqref{eq:proof_dF} to 
$p_q(w)=\operatorname{Pr}_m\{h(Z,w)\leq q\}=\operatorname{Pr}_m\{\mu(Z)+w\eta_m(Z)\leq q\}=\cF_{q}[\mu+w\eta_m]$, which gives \eqref{eq:dF}.

Now notice that
\begin{align*}
    \cF_{q}[\mum]-\cF_{q}[\mu] &= \cF_{q}[\mu+w\eta_m]\big|_{w=a_m}- \cF_{q}[\mu+w\eta_m]\big|_{w=0}\\
    &=\int_0^{a_m} \frac{d}{dw}\cF_{q}[\mu+w\eta_m] dw\\
    &=\int_0^{a_m} -\phi_{\mum}(q;w)\operatorname{E}[\eta_m(Z)|\mu(Z)+w\eta_m(Z)=q ]dw,
\end{align*}
and hence
\begin{align*}
    \left|\cF_{q}[\mum]-\cF_{q}[\mu]\right| &\leq
    \int_0^{a_m} \phi_{\mum}(q;w)dw \cdot \|\eta_m\|_\infty.
\end{align*}
Since
\begin{align*}
\int_0^{a_m} \phi_{\mum}(q;w)dw &= \int_0^{a_m} \int_{\mR} b_m(q-wu,u)dudw\\
&=\int_0^{a_m} \int_{\mR}\left[ b_m(q,u)- \partial_x b_m(q^*,u) \cdot w u\right] du dw\\
&\leq\int_0^{a_m} \int_{\mR}\left[ b_m(q,u) + |\partial_x b_m(q^*,u)| \cdot w |u|\right] du dw\\
&=\int_0^{a_m} \int_{\mR} b_m(q,u)du dw + \int_0^{a_m} \int_{\mR}|\partial_x b_m(q^*,u)| \cdot w |u| du dw\\
&\leq a_m\cdot f_{\mu}(q) + \frac{a_m^2}{2}\cdot B_{1,1}\\
&=\cO(a_m),
\end{align*}
where $q^\ast$ is between $q$ and $q-wu$.
Therefore $\cF_{q}[\mum]-\cF_{q}[\mu] = \cO(a_m) \cdot\|\eta_m\|_\infty=\cO(\|\mum-\mu\|_{\infty})= \cO_{\oP}(a_m)$.
\Halmos
\endproof

\subsection{Lemma~\ref{lem:g} and Its Proof}

\begin{customthm}{\ref{lem:g}}
Suppose Assumption~\ref{asmp:err_rate} holds. For sufficiently large $m$, $h_m(x,u,y,v)$, the joint density function  of $(\mu(Z),\eta_m(Z),\pi(Z),\gamma_m(Z))$, and its derivative $\partial_xh_m(x,u,y,v)$,  $\partial_yh_m(x,u,y,v)$ exist for all $x,u,y$ and $v$. In addition, let  $h_m^{Y}(x,u,y,v)=\int_{-\infty}^{y}h_m(x,u,\check{y},v)d\check{y}$. Its derivatives $\partial_xh_m^{Y}(x,u,y,v)$ and $\partial_x^2h_m^{Y}(x,u,y,v)$ exist for all $x,u,y$ and $v$. There exist non-negative functions $\bar h_{m,i}(u,v)$, $i=0,1,2$, and $\bar h^{Y}_{m,i}(u,v)$,  $i=0,1,2$, 
such that
\begin{align*}
    h_m(x,u,y,v)&\leq \bar h_{m,0}(u,v), \quad |\partial_xh_m(x,u,y,v)|\leq \bar h_{m,1}(u,v), \quad |\partial_yh_m(x,u,y,v)|\leq \bar h_{m,2}(u,v),\\
    h_m^{Y}(x,u,y,v)&\leq \bar h^{Y}_{m,0}(u,v), \quad |\partial_xh_m^{Y}(x,u,y,v)|\leq \bar h^{Y}_{m,1}(u,v), \quad |\partial_x^2h_m^{Y}(x,u,y,v)|\leq \bar h^{Y}_{m,2}(u,v)
\end{align*}
for all $x,u,y$ and $v$, with
\begin{align*}
    H_i&\equiv\sup_m\int_{\mR}\int_{\mR} (1+|u|+|v|)^r\bar h_{m,i}(u,v)dudv<\infty,\\
    H_i^{Y}&\equiv\sup_m\int_{\mR}\int_{\mR} (1+|u|+|v|)^r\bar h^{Y}_{m,i}(u,v)dudv<\infty
\end{align*}
for $i=0,1,2$ and $r=1, 2$. Then if $(x_m,y_m)\rightarrow(x,y)$ and $f_\mu(x)\neq 0$,  we have
\[
 {\partial_x}G_{\mum,\pim}(x_m,y_m)\rightarrow {\partial_x}G_{\mu,\pi}(x,y),\quad
 {\partial_y}G_{\mum,\pim}(x_m,y_m)\rightarrow {\partial_y}G_{\mu,\pi}(x,y).
\]
\end{customthm}
\proof{Proof of Lemma~\ref{lem:g}.} 
First of all, by noticing that $b_m(x,u)=\int_{\mR}\int_{\mR}h_m(x,u,y,v)dydv$, it is easy to check that the regularity conditions of Lemma~\ref{lem:f} hold by setting
\[
\bar{b}_{m,i}(u) = \int_{\mR} \bar h_{m,i}(u,v)dv
\]
for $i=0,1$. Therefore both the conclusions of Lemma~\ref{lem:f} and \ref{lem:F} hold.

Next, we consider the partial derivatives with respect to $y$. We start by representing $G_{\mu,\pi}$ and $G_{\mum,\pim}$ in terms of $h_m$ by
\begin{align*}
    G_{\mu,\pi}(x,y)&=\oPr\{\pi(Z)\le y|\mu(Z)=x\}\\
    &=\frac{\int_\mR\int_{-\infty}^y\int_\mR h_m(x,u,\check{y},v) du d\check{y} dv}{\int_\mR\int_\mR\int_\mR h_m(x,u,\check{y},v) du d\check{y} dv}\\
    &=\frac{\int_\mR\int_{-\infty}^y\int_\mR h_m(x,u,\check{y},v) du d\check{y} dv}{f_\mu(x)};\\
    G_{\mum,\pim}(x,y)&=\oPr_m\{\pi(Z)+a_m \gamma_m(Z)\le y|\mu(Z)+a_m\eta_m(Z)=x\}\\
    &=\frac{\int_\mR\int_{-\infty}^y\int_\mR h_m(x-a_m u,u,\check{y}-a_m v,v) du d\check{y} dv}{\int_\mR\int_\mR\int_\mR h_m(x-a_m u,u,\check{y}-a_m v,v) du d\check{y} dv}\\
    &=\frac{\int_\mR\int_{-\infty}^y\int_\mR h_m(x-a_m u,u,\check{y}-a_m v,v) du d\check{y} dv}{f_{\mum}(x)}.
\end{align*}
Since the integrations are interchangeable due to Fubini's theorem, we have
\begin{align}
    \partial_y G_{\mu,\pi}(x,y)
    &=\frac{\int_\mR\int_\mR h_m(x,u,y,v) du  dv}{f_\mu(x)};\label{eq:proof_Gy1}\\
    \partial_y G_{\mum,\pim}(x_m,y_m)
    &=\frac{\int_\mR\int_\mR h_m(x_m-a_m u,u,y_m-a_m v,v) du  dv}{f_{\mum}(x_m)}.\label{eq:proof_Gy2}
\end{align}
Notice that 
\begin{align*}
    & \left|\int_\mR\int_\mR h_m(x_m-a_m u,u,y_m-a_m v,v) du  dv - \int_\mR\int_\mR h_m(x,u,y,v) du  dv  \right|\\
    &= \left|\int_\mR\int_\mR\left[\partial_x  h_m(x^*,u,y^*,v) (x_m - a_m u -x) + \partial_y  h_m(x^*,u,y^*,v) (y_m - a_m v -y)\right]dudv\right|\\
    &\leq |x_m-x|\cdot\int_\mR\int_\mR |\partial_x  h_m(x^*,u,y^*,v)| dudv 
    + |y_m  -y|\cdot\int_\mR\int_\mR |\partial_y  h_m(x^*,u,y^*,v)| dudv \\
    &~~~~~~~~+a_m\left[\int_\mR\int_\mR |u|\cdot|\partial_x  h_m(x^*,u,y^*,v)| dudv+ \int_\mR\int_\mR |v|\cdot |\partial_y  h_m(x^*,u,y^*,v)| dudv\right]\\
    &\leq |x_m-x|\cdot\int_\mR\int_\mR \bar h_{m,1}(u,v) dudv 
    + |y_m  -y|\cdot\int_\mR\int_\mR \bar h_{m,2}(u,v) dudv \\
    &~~~~~~~~+a_m\left[\int_\mR\int_\mR |u|\cdot  \bar h_{m,1}(u,v) dudv+ \int_\mR\int_\mR |v|\cdot \bar  h_{m,}(u,v) dudv\right]\\
    &\leq |x_m-x|\cdot H_1 + |y_m  -y|\cdot H_2+a_m\cdot(H_1+H_2).
\end{align*}
As $x_m-x$, $y_m-y$ and $a_m$ converge to 0 when $m\rightarrow\infty$, the numerator of \eqref{eq:proof_Gy2} converges to that of \eqref{eq:proof_Gy1}. In addition, by Lemma~\ref{lem:f}, we know that the denominator of \eqref{eq:proof_Gy2} converges to that of \eqref{eq:proof_Gy1}. As a result, given that $f_\mu(x)\neq 0$, we have
${\partial_y}G_{\mum,\pim}(x_m,y_m)\rightarrow {\partial_y}G_{\mu,\pi}(x,y)$.

Last, we consider the partial derivatives with respect to $x$. We have
\begin{align*}
    \partial_x G_{\mu,\pi}(x,y)
    &=\frac{1}{f_\mu(x)}\int_\mR\int_{-\infty}^y\int_\mR \partial_xh_m(x,u,\check{y},v) du d\check{y} dv\\
    &~~~~~~~~-\frac{f'_\mu(x)}{[f_\mu(x)]^2}\int_\mR\int_{-\infty}^y\int_\mR h_m(x,u,\check{y},v) du d\check{y} dv\\
    &\equiv \frac{1}{f_\mu(x)} I_{m,1}(x,y)-\frac{f'_\mu(x)}{[f_\mu(x)]^2}I_{m,2}(x,y);\\
    \partial_x G_{\mum,\pim}(x,y)
    &=\frac{1}{f_{\mum}(x)}\int_\mR\int_{-\infty}^y\int_\mR \partial_xh_m(x-a_m u,u,\check{y}-a_m v,v) du d\check{y} dv\\
    &~~~~~~~~-\frac{f'_{\mum}(x)}{[f_{\mum}(x)]^2}\int_\mR\int_{-\infty}^y\int_\mR h_m(x-a_m u,u,\check{y}-a_m v,v) du d\check{y} dv\\
    &\equiv \frac{1}{f_{\mum}(x)} \tilde I_{m,1}(x,y)-\frac{f'_{\mum}(x)}{[f_{\mum}(x)]^2} \tilde I_{m,2}(x,y).
\end{align*}
By Lemma~\ref{lem:f}, we know that $f_{\mum}(x_m)\rightarrow f_{\mu}(x)$ and $f'_{\mum}(x_m)\rightarrow f_{\mu}(x)$. Then it suffices to show that $\tilde I_{m,i}(x_m,y_m)\rightarrow  I_{m,i}(x,y)$ for $i=1,2$. Actually,
\begin{align*}
    &|\tilde I_{m,1}(x_m,y_m) - I_{m,1}(x,y) |\\
    &= \left|\int_\mR\int_\mR\left[ \int_{-\infty}^{y_m}\partial_xh_m(x_m-a_m u,u,\check{y}-a_m v,v) d\check{y}-  \int_{-\infty}^y\partial_xh_m(x,u,\check{y},v)d\check{y} \right]du  dv\right|\\
    &= \left|\int_\mR\int_\mR\left[ \int_{-\infty}^{y_m-a_mv}\partial_xh_m(x_m-a_m u,u,\check{y},v) d\check{y}-  \int_{-\infty}^y\partial_xh_m(x,u,\check{y},v)d\check{y} \right]du  dv\right|\\
    &= \left|\int_\mR\int_\mR\left[ \partial_xh_m^{Y}(x_m-a_m u,u,y_m-a_mv,v)-  \partial_xh_m^{Y}(x,u,y,v) \right]du  dv\right|\\
    &= \left|\int_\mR\int_\mR\left[ (x_m-a_mu-x)\partial^2_xh_m^{Y}(x^{**},u,y^{**},v) + (y_m-a_mv-y)\partial_xh_m(x^{**},u,y^{**},v)    \right]du  dv\right|\\
    &\leq |x_m-x|\cdot \int_\mR\int_\mR \bar h^{Y}_{m,2}(u,v)dudv + |y_m-y|\cdot \int_\mR\int_\mR \bar h_{m,1}(u,v)dudv \\
    &~~~~~~~~+a_m\left[\int_\mR\int_\mR |u|\bar h^{Y}_{m,2}(u,v)dudv+\int_\mR\int_\mR |v|\bar h_{m,1}(u,v)dudv\right]\\
     &\leq |x_m-x|\cdot H_2^{Y}+|y_m-y|\cdot H_1 + a_m\cdot(H_2^{Y}+H_1),
\end{align*}
and
\begin{align*}
    &|\tilde I_{m,2}(x_m,y_m)-I_{m,2}(x,y)|\\
    &= \left|\int_\mR\int_\mR\left[ \int_{-\infty}^{y_m}h_m(x_m-a_m u,u,\check{y}-a_m v,v) d\check{y}-  \int_{-\infty}^yh_m(x,u,\check{y},v)d\check{y} \right]du  dv\right|\\
    &= \left|\int_\mR\int_\mR\left[ (x_m-a_mu-x)\partial_xh_m^{Y}(x^{***},u,y^{***},v)  + (y_m-a_mv-y)h_m(x^{***},u,y^{***},v)    \right]du  dv\right|\\
    &\leq |x_m-x|\cdot \int_\mR\int_\mR \bar h^{Y}_{m,1}(u,v)dudv + |y_m-y|\cdot \int_\mR\int_\mR \bar h_{m,0}(u,v)dudv \\
    &~~~~~~~~+a_m\left[\int_\mR\int_\mR |u|\bar h^{Y}_{m,1}(u,v)dudv+\int_\mR\int_\mR |v|\bar h_{m,0}(u,v)dudv\right]\\
     &\leq |x_m-x|\cdot  H_1^{Y}+|y_m-y|\cdot H_0 + a_m\cdot(H_1^{Y}+H_0).
\end{align*}
As $x_m-x$, $y_m-y$ and $a_m$ converge to 0 when $m\rightarrow\infty$, we know that $\tilde I_{m,i}(x_m,y_m)\rightarrow  I_{m,i}(x,y)$ for $i=1,2$.
Given that $f_{\mu}(x)\neq 0$, we have
${\partial_x}G_{\mum,\pim}(x_m,y_m)\rightarrow {\partial_x}G_{\mu,\pi}(x,y)$.
\Halmos
\endproof

\subsection{Lemma~\ref{lem:G} and Its Proof}
\begin{alemma}\label{lem:df_cont_a}
Denote $\Psi_m(x,y;w,s)$ and $\psi_m(x,y;w,s)$ be the joint cumulative distribution and density function of $(\mu(Z)+w\eta_m(Z), \pi(Z)+s\gamma_m(Z))$ for $w,s\in[0,a_m]$, respectively. Recall that $\Phi_m(x;w)$ and $\phi_m(x;w)$ are the marginal cumulative distribution function and density of $\mu(Z)+w\eta_m(Z)$, respectively. Assume Assumption~\ref{asmp:err_rate} and regularity conditions of Lemma~\ref{lem:g} hold. In addition, let $h_m^{X}(x,u,y,v)=\int_{-\infty}^{x}h_m(\check{x},u,y,v)d\check{x}$ and its derivatives $\partial_yh_m^{X}(x,u,y,v)$ exist for all $x,u,y$ and $v$. There exist non-negative functions $\bar h^{X}_{m,i}(u,v)$  $i=0,1$ such that
\begin{align*}
    h_m^{X}(x,u,y,v)&\leq \bar h^{X}_{m,0}(u,v), \quad |\partial_yh_m^{X}(x,u,y,v)|\leq \bar h^{X}_{m,1}(u,v),
\end{align*}
for all $x,u,y$ and $v$, with
\begin{align*}
    H_i^{X}&\equiv\sup_m\int_{\mR}\int_{\mR} (1+|u|+|v|)\bar h^{X}_{m,i}(u,v)dudv<\infty,
\end{align*}
for $i=0,1,2$. Then we have
\begin{itemize}
    \item[(a)] For all $w,s\in(0,a_m)$, $\partial_w \Psi_m(x, y; w, s)$, $\partial_x^2 \Psi_m(x, y; w, s)$, and $\partial_w \partial_x \Psi_m(x, y; w, s)$ exist and are continuous with respect to $w\in(0,a_m)$, $x$ in a neighbourhood of $q$, and $y$ in a neighbourhood of $\theta$. Moreover, $\partial_w \Phi_m(x; w), \partial_x \phi_m(x; w), \partial_w \phi_m(x; w)$ exist and are continuous with respect to both $w\in(0,a_m)$ and $x$ in a neighbourhood of $q$. For all $w\in(0,a_m)$, $\phi_m\left(q, w\right)>0$.

    \item[(b)] For all $w,s\in(0,a_m)$, $\partial_x \Psi_m(x, y; w, s)$, $\partial_s \Psi_m(x, y; w, s)$, and $\partial_s \partial_x \Psi_m(x, y; w, s)$ exist and are continuous with respect to $s\in(0,a_m)$, $x$ in a neighbourhood of $q$, and $y$ in a neighbourhood of $\theta$.
\end{itemize}
\end{alemma}
\proof{Proof of Lemma~\ref{lem:df_cont_a}.}
First of all, by definition, we have
\begin{align*}
\Psi_m(x,y;w,s)&=\int_\mR\int_\mR \int_{-\infty}^{x-w u}\int_{-\infty}^{y-s v} h_m(\check{x},u,\check{y},v) d\check{y}  d\check{x} du dv, \\
\psi_m(x,y;w,s)&=\int_\mR\int_\mR h_m(x-w u,u,y-s v,v) du dv,\\
\Phi_m(x;w)&=\lim_{y\rightarrow\infty}\Psi_m(x,y;w,s)=\int_{\mR}\int_\mR\int_\mR\int_{-\infty}^x h_m(\check{x}-w u,u,y,v) d\check{x}du dvdy,\\
\phi_m(x;w)&=\int_{\mR}\psi_m(x,y;w,s)dy=\int_\mR\int_\mR\int_\mR h_m(x-w u,u,y,v) du dv dy.
\end{align*}
Therefore, we have
\begin{align*}
   \partial_w\Psi_m(x,y;w,s)&=-\int_\mR\int_\mR \int_{-\infty}^{y-s v} u
   \cdot h_m(x-wu,u,\check{y},v) d\check{y}   du dv\\
   &=-\int_\mR\int_\mR  u
   \cdot h_m^{Y}(x-wu,u,y-s v,v)   du dv,\\
   \partial_x\Psi_m(x, y; w, s)&=\int_\mR\int_\mR \int_{-\infty}^{y-s v} 
   h_m(x-wu,u,\check{y},v) d\check{y}   du dv\\
   &=\int_\mR\int_\mR 
   h_m^{Y}(x-wu,u,y-sv,v) du dv,\\
   \partial_x^2 \Psi_m(x, y; w, s)
   &=\int_\mR\int_\mR 
   \partial_xh_m^{Y}(x-wu,u,y-sv,v) du dv,\\
    \partial_w\partial_x \Psi_m(x, y; w, s)
   &=-\int_\mR\int_\mR 
 u\cdot  \partial_xh_m^{Y}(x-wu,u,y-sv,v) du dv,\\
 \partial_s\Psi_m(x,y;w,s)&=-\int_\mR\int_\mR \int_{-\infty}^{x-wu} v
   \cdot h_m(\check{x},u,y-sv,v) d\check{x}   du dv\\
   &=-\int_\mR\int_\mR  v
   \cdot h_m^{X}(x-wu,u,y-s v,v)   du dv,\\
   \partial_s\partial_x \Psi_m(x, y; w, s)
   &=-\int_\mR\int_\mR v
   \cdot h_m(x-wu,u,y-sv,v)   du dv.\\
    \partial_w \Phi_m(x; w)&=-\int_\mR\int_\mR  u
   \cdot h_m^{Y}(x-wu,u,\infty,v)   du dv,\\
    \partial_x \phi_m(x; w)&=\int_\mR\int_\mR \partial_xh_m^{Y}(x-w u,u,\infty,v) du dv,\\
    \partial_w \phi_m(x; w)&=-\int_\mR\int_\mR u\cdot  \partial_xh_m^{Y}(x-w u,u,\infty,v) du dv.
\end{align*}
Then the continuity of these functions follows from the regularity conditions in Lemma~\ref{lem:g} and Lemma~\ref{lem:df_cont_a}.
\Halmos
\endproof

\begin{alemma}[\citealt{fu2023sensitivity}]\label{lem:fu2003}
Suppose $X(w)$ and $Y(s)$ are two random variables with parameters $w\in\Theta$, $s\in\Delta$, where $\Theta$ and $\Delta$ are open subset of $\mR$.
Let $f_{X,Y}(x, y; w, s)$ and $F_{X, Y}(x, y; w, s)$ be the joint probability density function and cumulative distribution function of $(X(w), Y(s))$, respectively, and $f_X(x; w)$ and $F_X(x; w)$ be the marginal probability density function and cumulative distribution function of $X(w)$, respectively.
If the following conditions are satisfied:

{\rm (a)} The pathwise derivative $\partial_w X(w), \partial_s Y(s)$ exist w.p.1 for any $w \in \Theta$ and $s \in \Delta$, and there exist random variables $K$ and $T$ with $\oE[K],\oE[T]<\infty$, which are not dependent on $w$ and $s$, such that $\left|X\left(w_1\right)-X\left(w_2\right)\right| \leq$ $K\left|w_1-w_2\right|$ and $\left|Y\left(s_1\right)-Y\left(s_2\right)\right| \leq T\left|s_1-s_2\right|$ hold for all $w_1, w_2 \in \Theta, s_1, s_2 \in \Delta$.

{\rm (b)} For all $w \in \Theta, s \in \Delta, \partial_w F_{X, Y}(x, y; w, s)$, $\oE\left[I_{\{X(w) \leq x\}} I_{\{Y(s) \leq y\}} \partial_w X(w)\right], \partial_x^2 F_{X, Y}(x, y; w, s)$, and $\partial_w \partial_x F_{X, Y}(x, y; w, s)$ exist and are continuous with respect to $w\in\Theta$, $x$ in a neighbourhood of $a$, and $y$ in a neighbourhood of $b$. Moreover, $\partial_w F_X(x; w), \partial_x f_X(x; w), \partial_w f_X(x; w), $ and $\oE\left[I_{\{X(w) \leq x\}} \partial_w X(w)\right]$ exist and are continuous with respect to both $w\in\Theta$ and $x$ in a neighbourhood of $a$. For all $w\in\Theta$, $f_X\left(a, w\right)>0$.

{\rm (c)} For all $w \in \Theta, s \in \Delta, \partial_x F_{X, Y}(x, y; w, s)$, $\oE\left[I_{\{X(w) \leq x\}} I_{\{Y(s) \leq y\}} \partial_s Y(s)\right]$, $\partial_s F_{X, Y}(x, y; w, s)$, and $\partial_s \partial_x F_{X, Y}(x, y; w, s)$ exist and are continuous with respect to $s\in\Delta$, $x$ in a neighbourhood of $a$, and $y$ in a neighbourhood of $b$.

Then for $p_{a,b}(w,s)=\Pr\{Y(s)\leq b|X(w)=a\}$, we have
$$
\partial_w p_{a, b}(w, s)=\frac{\left.\left\{-\partial_x^2 \oE\left[I_{\{X(w) \leq x\}} I_{\{Y(s) \leq b\}} \partial_w X(w)\right]+p_{a, b}(w, s) \partial_x^2 \oE\left[I_{\{X(w) \leq x\}} \partial_w X(w)\right]\right\}\right|_{x=a}}{f_X(a, w)}
$$
$$
\partial_s p_{a, b}(w, s)=\frac{-\left.\partial_x \partial_y \oE\left[I_{\{X(w) \leq x\}} I_{\{Y(s) \leq y\}} \partial_s Y(s)\right]\right|_{x=a, y=b}}{f_X(a, w)}
$$
\end{alemma}

\begin{customthm}{\ref{lem:G}}
Under Assumptions~\ref{asmp:err_rate} and regularity conditions of Lemma~\ref{lem:g} and Lemma~\ref{lem:df_cont_a}, we have
\begin{equation*}
    \cG_{q,\theta}[\mum,\pim]-\cG_{q,\theta}[\mu,\pi] = \cO_{\oP}(a_m).
\end{equation*}
\end{customthm}

\proof{Proof of Lemma~\ref{lem:G}.}
Define $p_{q,\theta}(w,s)=\Pr\{\pi(Z)+s\gamma_m(Z)\leq \theta|\mu(Z)+w\eta_m(Z)=q\}$, then
\[
\cG_{q,\theta}[\mum,\pim]= p_{q,\theta}(a_m,a_m),\quad 
\cG_{q,\theta}[\mu,\pi]= p_{q,\theta}(0,0).
\]
Under regularity conditions of Lemma~\ref{lem:g}, it is straightforward to check the conditions in Lemma~\ref{lem:fu2003} hold with $X(w)=\mu(Z)+w\eta_m(Z)$, $Y(s)=\pi(Z)+s\gamma_m(Z)$, $a=q$ and $b=\theta$. Consequently, we can apply the conclusions of Lemma~\ref{lem:fu2003} to derive partial derivatives of $p_{q,\theta}(w,s)$. In particular, with $\partial_w X(w)=\eta_m(Z)$, $\partial_s Y(s)=\gamma_m(Z)$, we have
\begin{align}
  \partial_w p_{q, \theta}(w, s)&=\frac{\left.\left\{-\partial_x^2 \oE\left[I_{\{X(w) \leq x\}} I_{\{Y(s) \leq \theta\}} \eta_m(Z)\right]+p_{q, \theta}(w, s) \partial_x^2 \oE\left[I_{\{X(w) \leq x\}} \eta_m(Z)\right]\right\}\right|_{x=q}}{\phi_m(q; w)},\label{eq:proof_pd1}\\
  \partial_s p_{q, \theta}(w, s)&=\frac{-\left.\partial_x \partial_y \oE\left[I_{\{X(w) \leq x\}} I_{\{Y(s) \leq y\}} \gamma_m(Z)\right]\right|_{x=q, y=\theta}}{\phi_m(q; w)}.\label{eq:proof_pd2}
\end{align}
Under regularity conditions of Lemma~\ref{lem:g}, the partial derivatives and expectations are interchangeable, and thus
\begin{align*}
    \partial_x \oE\left[I_{\{X(w) \leq x\}} I_{\{Y(s) \leq y\}} \eta_m(Z)\right]&= \partial_x \int_{-\infty}^x\oE\left[I_{\{Y(s) \leq y\}} \eta_m(Z)|X(w)=t\right]\phi_m(t;w)dt\\
    &=\oE\left[I_{\{Y(s) \leq y\}} \eta_m(Z)|X(w)=x\right]\phi_m(x;w),
\end{align*}
and similarly
\begin{align*}
    \partial_x \oE\left[I_{\{X(w) \leq x\}} \eta_m(Z)\right]
    &=\oE\left[\eta_m(Z)|X(w)=x\right]\phi_m(x;w),\\
    \partial_x \oE\left[I_{\{X(w) \leq x\}} I_{\{Y(s) \leq y\}} \gamma_m(Z)\right]
    &=\oE\left[I_{\{Y(s) \leq y\}} \gamma_m(Z)|X(w)=x\right]\phi_m(x;w).
\end{align*}
Furthermore, under regularity conditions of Lemma~\ref{lem:g}, the partial derivatives $\partial_x$ and $\partial_y$ are also interchangeable, and thus
\begin{align*}
    \partial_x \partial_y \oE\left[I_{\{X(w) \leq x\}} I_{\{Y(s) \leq y\}} \gamma_m(Z)\right]
    &=\partial_y\left\{\oE\left[I_{\{Y(s) \leq y\}} \gamma_m(Z)|X(w)=x\right]\phi_m(x;w)\right\}\\
    &=\phi_m(x;w)\partial_y\int_{-\infty}^y\oE\left[\gamma_m(Z)|X(w)=x,Y(s)=\check y\right] f_{Y|X}(\check y|x)d \check y\\
    &=\oE\left[\gamma_m(Z)|X(w)=x,Y(s)=y\right] \psi_m(x,y;w,s),
\end{align*}
which indicates
\begin{align}\label{eq:proof_dd1}
    \left|\partial_x \partial_y \oE\left[I_{\{X(w) \leq x\}} I_{\{Y(s) \leq y\}} \gamma_m(Z)\right]\right|
    &\leq  \|\gamma_m\|_{\infty} \psi_m(x,y;w,s)=   \cO_{\oP}(1).
\end{align}

We also have
\begin{align*}
    \partial_x^2\oE\left[I_{\{X(w) \leq x\}} \eta_m(Z)\right]&=\partial_x\left\{\oE\left[\eta_m(Z)|X(w)=x\right]\phi_m(x;w)\right\}\\
    &=\partial_x \int_{\mR} u b_m(x-wu,u)du\\
    &=\int_{\mR} u \partial_x b_m(x-wu,u)du.
\end{align*}
Due to the regularity conditions, we have
\begin{align}\label{eq:proof_dd2}
   \left| \partial_x^2\oE\left[I_{\{X(w) \leq x\}} \eta_m(Z)\right]\right|
    &\leq\int _\mR |u|  \bar{b}_{m,1}(u)du \leq B_{1,1}.
\end{align}

Similarly,
\begin{align*}
    \partial_x^2\oE\left[I_{\{X(w) \leq x\}}I_{\{Y(s) \leq y\}}  \eta_m(Z)\right]&=\partial_x\left\{\oE\left[I_{\{Y(s) \leq \theta\}} \eta_m(Z)|X(w)=x\right]\phi_m(x;w)\right\}\\
    &=\partial_x \int_{-\infty}^{y-sv}\int_{\mR} \int_{\mR}u h_m(x-wu,u,\check{y},v)dudvd\check{y}\\
    &=\int_{\mR} \int_{\mR}u \partial_x h_m^{Y}(x-wu,u,y-sv,v)dudv.
\end{align*}
Due to the regularity conditions, we have
\begin{align}\label{eq:proof_dd3}
   \left| \partial_x^2 \oE\left[I_{\{X(w) \leq x\}} I_{\{Y(s) \leq \theta\}} \eta_m(Z)\right]\right|
    &\leq\int_{\mR}\int_{\mR} |u| \bar{h}_{m,1}(u,v) dudv\leq H_1^{Y}.
\end{align}

Substitute \eqref{eq:proof_dd1}, \eqref{eq:proof_dd2} and \eqref{eq:proof_dd3} into \eqref{eq:proof_pd1} and \eqref{eq:proof_pd2}, we known both $\partial_w p_{q, \theta}(w, s)$ and $\partial_w p_{q, \theta}(w, s)$ are of order $\cO_{\oP}(1)$ as $m\rightarrow\infty$. Now define $g(t)=\cG_{q,\theta}[\mu+t\eta_m, \pi+t \gamma_m]$, $t\in[0,a_m]$. Notice that $g(t)=p_{q,\theta}(t,t)$. Therefore
\begin{align*}
    g'(s)=\partial_w p_{a, b}(s, s) + \partial_s p_{a, b}(s, s)=\cO_{\oP}(1).
\end{align*}
Therefore
\begin{align*}
\cG_{q,\theta}[\mum, \pim] -\cG_{q,\theta}[\mu, \pi]=g(a_m)-g(0)=\int_0^{a_m} g'(s)ds = \cO_{\oP}(a_m).
\end{align*}
\Halmos
\endproof

\subsection{Lemma~\ref{lem:bi-taylor} and Its Proof}
\begin{alemma}\label{lem:bi-taylor}
(a) Under the regularity conditions of Lemma~\ref{lem:f}, we have
\begin{align*}
F_{\mum}(\qm)=F_{\mum}(q)+f_{\mum}(q)(\qm-q)+o(\qm-q).
\end{align*}
(b) Under the regularity conditions of Lemma~\ref{lem:g}, we have
\begin{align*}
     G_{\mum,\pim}(\qm,\tm)-G_{\mum,\pim}(q,\theta)&=\partial_xG_{\mum,\pim}(q^*,\theta)(\qm-q)+\partial_yG_{\mum,\pim}(q,\theta)(\tm-\theta)+o(\tm-\theta),
\end{align*}
as $m\to \infty$, where $q^*$ is some number between $\qm$ and $q$.
\end{alemma}
\proof{Proof of Lemma~\ref{lem:bi-taylor}.}
(a) By the mean value theorem, we have
\[
F_{\mum}(\qm)=F_{\mum}(q)+f_{\mum}(q)(\qm-q)+\frac{1}{2}f'_{\mum}(q^*)(\qm-q)^2
\]
for some $q^*$ between $q$ and $\qm$. Following the proof of Lemma~\ref{lem:f}, we can derive $f'_{\mum}(x)$ by switching integration and differentiation, which yields
\begin{align*}
    f_{\mum}'(x) &=\frac{d}{dx}\int_{\mR}b_m(x-a_m u,u)du= \int_{\mR}\partial_xb_m(x-a_m u,u)du.
\end{align*}
Hence, under the regularity conditions of Lemma~\ref{lem:f},
    $|f_{\mum}'(x)|\leq \int_{\mR}|\partial_xb_m(x-a_m u,u)|du\leq \int_{\mR}\bar b_{m,1}(u)du\leq B_{0,1}$. As a result,
\[
F_{\mum}(\qm)=F_{\mum}(q)+f_{\mum}(q)(\qm-q)+o(\qm-q).
\]

(b) By the mean value theorem, we have
\begin{align*}
    G_{\mum,\pim}(\qm,\tm)-G_{\mum,\pim}(q,\tm)&=\partial_xG_{\mum,\pim}(q^*,\tm)(\qm-q),\\
    G_{\mum,\pim}(q,\tm)-G_{\mum,\pim}(q,\theta)&=\partial_yG_{\mum,\pim}(q,\theta)(\tm-\theta)+\frac{1}{2}\partial_{yy}G_{\mum,\pim}(q,\theta^*)(\tm-\theta)^2,\\
  \partial_xG_{\mum,\pim}(q^*,\tm)&=\partial_xG_{\mum,\pim}(q^*,\theta)+\partial_{xy}G_{\mum,\pim}(q^{*},\theta^{**})(\tm-\theta),
\end{align*}
where $\theta^*, \theta^{**}$ are between $\tm$ and $\theta$, and $q^*$ is between $\qm$ and $q$. Therefore we have
\begin{align*}
     G_{\mum,\pim}(\qm,\tm)-G_{\mum,\pim}(q,\theta)&=\partial_xG_{\mum,\pim}(q^*,\theta)(\qm-q)+\partial_yG_{\mum,\pim}(q,\theta)(\tm-\theta)+R_m.
\end{align*}
where $R_m = \partial_{xy}G_{\mum,\pim}(q^{*},\theta^{**})(\qm-q)(\tm-\theta)+\frac{1}{2}\partial_{yy}G_{\mum,\pim}(q,\theta^*)(\tm-\theta)^2$. Therefore, it suffice to prove that $R_m=o(\tm-\theta)$ as $m\to\infty$.

Following equation~\eqref{eq:proof_Gy2} in the proof of Lemma~\ref{lem:g},
\begin{align*}
    \partial_y G_{\mum,\pim}(x,y)
    &=\frac{\int_\mR\int_\mR h_m(x-a_m u,u,y-a_m v,v) du  dv}{f_{\mum}(x)}.
\end{align*}
By further taking derivative with respect to $y$, we have
\begin{align*}
    \partial^2_y G_{\mum,\pim}(x,y)
    &=\frac{\int_\mR\int_\mR \partial_yh_m(x-a_m u,u,y-a_m v,v) du  dv}{f_{\mum}(x)},
\end{align*}
and hence
\begin{align*}
    \left|\partial^2_y G_{\mum,\pim}(q,\theta^*)\right|
    &\leq \frac{\int_\mR\int_\mR \bar h_{m,2}(u,v) du  dv}{f_{\mum}(q)}
    \leq \frac{H_2}{f_{\mum}(q)} \to \frac{H_2}{f_{\mu}(q)}.
\end{align*}
Similarly, by further taking derivative with respect to $x$, we have
\begin{align*}
\partial^2_{xy} G_{\mum,\pim}(x,y)
    &=\frac{1}{f_{\mum}(x)}\int_\mR\int_\mR \partial_xh_m(x-a_m u,u,y-a_m v,v) du  dv\\
    &~~~~~~~~-\frac{f'_{\mum}(x)}{[f_{\mum}(x)]^2}\int_\mR\int_\mR h_m(x-a_m u,u,y-a_m v,v) du dv,
\end{align*}
and hence
\begin{align*}
\left|\partial^2_{xy} G_{\mum,\pim}(q^*,\theta^{**})\right|
    &\leq \frac{1}{f_{\mum}(q^*)}\int_\mR\int_\mR \bar h_{m,1}(u,v) du  dv+\frac{|f'_{\mum}(q^*)|}{[f_{\mum}(q^*)]^2}\int_\mR\int_\mR \bar h_{m,0}(u,v) du dv\\
    &\leq\frac{1}{f_{\mum}(q^*)}H_1+\frac{|f'_{\mum}(q^*)|}{[f_{\mum}(q^*)]^2}H_0\\
    &\to\frac{1}{f_{\mu}(q)}H_1+\frac{|f'_{\mu}(q)|}{[f_{\mu}(q)]^2}H_0,
\end{align*}
where for the last limit we have used Lemma~\ref{lem:f} and Proposition~\ref{prop:qm}. Combine the above results with the definition of $R_m$, we can conclude that $R_m=o(\tm-\theta)$.
\Halmos
\endproof

\section{Other Proofs}\label{sec:other-proofs}


\proof{Proof Sketch of Proposition \ref{prop:CoVaR_SNS}.}\label{proof:CoVaR_SNS}
Given that the complete proof of this proposition is both tedious and standard, adhering closely to established methods in the literature, we provide only a sketch of the proof to highlight the key steps and underlying logic.
Let $\xi_j(Z)$ and $\zeta_j(Z)$ denote the portfolio pricing error for the $j-$th inner-level observation of $\mu(z)$ and $\pi(z)$ respectively, and let $\bar{\xi}^l(Z)= {1\over l}\sum_{j=1}^l \xi_j(Z)$ and $\bar{\zeta}^l(Z)= {1\over l}\sum_{j=1}^l\zeta_j(Z)$
be the zero-mean average pricing errors for the portfolios. Following the analytical scheme of \cite{gordy2010nested}, we take the surrogates
$\hat{X}(Z)=\mu(Z)+\bar{\xi}^l(Z)$ and $\hat{Y}(Z)=\pi(Z)+\bar{\zeta}^l(Z)$ and analyze them instead of analyzing $\bar{X}$ and $\bar{Y}$ directly.
Let $\hat \xi^l = \sqrt{l}\bar{\xi}^l$ and $\hat \zeta^l = \sqrt{l}\bar{\zeta}^l$ so that they have nontrivial limit distribution as $l\rightarrow \infty$. 
Let $G_{\mu,\pi}(x,y)=\operatorname{Pr}\{\pi(Z)\leq y|\mu(Z)=x\}$ and $G_{\hat X,\hat Y}(x,y)=\operatorname{Pr}\{\hat Y(Z)\leq y|\hat X(Z)=x\}$. Define the joint density of $(\mu,\pi,\hat{\xi}_l, \hat{\zeta}_l)$ as $\phi(s,t,u,v)$. With the regularity conditions on the joint density and its partial derivatives similar to Assumption 1 in \cite{gordy2010nested}, we have 
\begin{equation}\label{eq:sns_func}
G_{\hat X, \hat Y}(x,y)-G_{\mu, \pi}(x,y)=\cO({1/l}).
\end{equation}

Let $\hat{\theta}_{k,h,l}$ be the CoVaR estimator of standard nested estimator  and $\theta$ be the $(\alpha,\beta)$-CoVaR  of $(\mu(Z),\pi(Z))$, which satisfies $\oP\{\pi(Z)\leq \theta|\mu(Z)=q_\alpha\}=G_{\mu,\pi}(q_\alpha,\theta)=\beta$, where $q_\alpha$ is the $\alpha$-VaR of $\mu(Z)$.
To derive the mean squared error (MSE) $\oE[(\hat{\theta}_{k,h,l}-\theta)^2]$, we introduce another two auxiliary variables. We define $\hat \theta_{h,l}$ such that $\oP\{\hat{Y}(Z)\leq \hat{\theta}_{h,l}|\hat{X}(Z)=\bar{X}_{\lceil h\alpha\rceil}\}=G_{\hat X, \hat{Y}}(\bar{X}_{\lceil h\alpha\rceil},\hat{\theta}_{h,l})=\beta$, where $\bar{X}_{\lceil h\alpha\rceil}$ is the ${\lceil h\alpha\rceil}$-th order statistics of $\bar{X}$. Besides, $\hat \theta_l$ is defined as the $(\alpha,\beta)$-CoVaR  of $(\hat X(Z),\hat Y(Z))$, which satisfies $\oP\{\hat Y(Z)\leq \hat\theta_l|\hat X(Z)=\hat q\}=G_{\hat X, \hat{Y}}(\hat q,\hat{\theta}_{l})=\beta$, where $\hat q$ is the $\alpha$-VaR of $\hat X(Z)$. 

The MSE, $\oE[(\hat{\theta}_{k,h,l}-\theta)^2]$, can be decomposed into 
\[
\oE[(\hat{\theta}_{k,h,l}-\theta)^2]
= \underbrace{{\rm VaR}(\hat{\theta}_{k,h,l})}_{variance} + \underbrace{(\oE[\hat{\theta}_{k,h,l}]-\theta)^2}_{bias^2}.
\]
It's obvious that the order of variance term is $\cO(1/k)$ since it involves estimating the quantile with $k$ i.i.d observations.
Then, we analyze the bias term by further decompose the bias into:
\[
\oE[\hat{\theta}_{k,h,l}]-\theta
= \underbrace{\oE[\oE[\hat{\theta}_{k,h,l}|\bar{X}_{\lceil h\alpha\rceil}]-\hat{\theta}_{h,l}]}_{I}
+\underbrace{\oE[\hat{\theta}_{h,l}-\hat{\theta}_l]}_{II}
+ \underbrace{\hat{\theta}_l-\theta}_{III},
\]
and analyze them term by term.

For term I, $\oE[\hat{\theta}_{k,h,l}|\bar{X}_{\lceil h\alpha\rceil}]-\hat{\theta}_{h,l}]$ is intrinsically the difference between the $\lceil k\beta\rceil$-th order statistics and the $\beta$-th quantile of $\hat{Y}|\hat{X}=\bar{X}_{\lceil h\alpha\rceil}$.
Under the regularity conditions on conditional density of $\hat{Y}|\hat{X}$ and the marginal density of $\hat{X}$, we can obtain that $\oE[\oE[\hat{\theta}_{k,h,l}|\bar{X}_{\lceil h\alpha\rceil}]-\hat{\theta}_{h,l}]]=\cO(1/k)$ according to Equation (4.6.3) of \cite{david2004order}.

For term II, we apply Taylor's expansion to $G_{\hat{X},\hat{Y}}(\hat{\theta}_{h,l},\bar{X}_{\lceil h\alpha\rceil})$ at $(\hat{\theta}_l,\hat{q})$, we obtain that $\hat{\theta}_{h,l}-\hat{\theta}_l = \cO(\bar{X}_{\lceil h\alpha\rceil}-\hat{q})$ and thus $\oE[\hat{\theta}_{h,l}-\hat{\theta}_l] = \cO(\oE[\bar{X}_{\lceil h\alpha\rceil}-\hat{q}])=\cO(1/h)$ according to Lemma 2 of \cite{hong2009estimating}.

For term III, by applying Taylor's expansion to $G_{\hat{X},\hat{Y}}(\hat{\theta}_l,\hat{q})$ on $(\theta,q_\alpha)$ and using the fact that $G_{\hat{X},\hat{Y}}(\hat{\theta}_l,\hat{q})=G_{\mu,\pi}(q_\alpha,\theta)=\beta$, we can derive that
\[
\hat{\theta}_l-\theta = \cO(\underbrace{G_{\hat{X},\hat{Y}}(q_\alpha,\theta)-G_{\mu,\pi}(q_\alpha,\theta)}_{(i)})+\cO(\underbrace{\hat{q}-q}_{(ii)}),
\]
where term (i) is of order $\cO(1/l)$ by Equation \eqref{eq:sns_func} and term (i) is of order $\cO(1/l)$ by Proposition 2 of \cite{gordy2010nested}.

Therefore, the bias term is of order $\cO(1/k+1/h+1/l)$. 
Overall, the order of MSE is $\cO(1/k+1/h^2+1/l^2)$. 
Notice that the first two terms  of this rate, $1/k$ and $1/h^2$, align with the results derived in \cite{huang2022monte} under the assumption that the closed form expressions of $\mu(Z)$ and $\pi(Z)$ are available, while the third therm, $1/l^2$, arises from the inner-level estimation with $l$ sample observations.
Since $\Gamma = khl$, we have the best convergence rate of RMSE is $\cO(\Gamma^{-1/4})$, with $l = c_1\Gamma^{1/4}$, $h = c_2\Gamma^{1/4}$ and $k = {1\over c_1c_2}\Gamma^{1/2}$, for some constants $c_1, c_2>0$.
\Halmos
\endproof
\proof{Proof of Proposition~\ref{prop:smoothness}.}
We can rewrite $\mu(z)$ by
\begin{align*}
\mu(z)&=\oE[\varphi(Z)|Z_0=z]\\
&=\oE[\varphi(\psi_T\left(S_1, S_2, \cdots, S_n\right))|\psi_{0}(S_0)=z]\\
&= \iint\cdots\int \varphi(\psi_T\left(s_{1}, s_{2} \cdots, s_{n}\right))p_{0,1}(\psi_0^{-1}(z),s_1)\prod_{k=2}^n p_{{k-1},{k}}(s_{k-1},s_{k}) {\rm d} s_{1}{\rm d} s_{2}\cdots{\rm d} s_{n}.
\end{align*}
Let $\nu(s)=\mu(\psi_0(s))$, then $\mu(z)=\nu(\psi_0^{-1}(z))$. Since $\psi_0^{-1}$ is smooth, we only need to prove the smoothness of $\nu(\cdot)$. Actually, under condition (a) we can apply Fubini's theorem, and
\begin{align*}
    \nu(s) &= \iint\cdots\int \varphi(\psi_T\left(s_{1}, s_{2} \cdots, s_{n}\right))p_{0,1}(s,s_1)\prod_{k=2}^n p_{{k-1},{k}}(s_{k-1},s_{k}) {\rm d} s_{1}{\rm d} s_{2}\cdots{\rm d} s_{n} \\
    &= \int p_{0,1}(s,s_1) \left[\int\cdots\int \varphi(\psi_T\left(s_{1}, s_{2} \cdots, s_{n}\right))\prod_{k=2}^n p_{{k-1},{k}}(s_{k-1},s_{k}) {\rm d} s_{2}\cdots{\rm d} s_{n}\right]{\rm d} s_{1}\\
    &=\int p_{0,1}(s,s_1) \phi(s_1) {\rm d} s_{1}.
\end{align*}
For given $s$ and $j\in\{1,\cdots,d\}$, define a mapping $\zeta_{s,j}:\mR\rightarrow\mR^q$, such that $\zeta_j(t)$ has the $j$-th entry equal to $t\in\mR$ while all the other entries identical to the corresponding entries of $s$. Denote $e_j$ to be a $q$-dimensional vector with the $j$th entry being 1 and all the other entries being 0, and $s(j)$ to be the $j$-th entry of $s$, then by applying Leibniz integral rule,
\[
p_{0,1}(s,s_1) = p_{0,1}(\zeta_{s,j}(s(j)),s_1)=p_{0,1}(\zeta_{s,j}(t_0),s_1)+\int_{t_0}^{s(j)} \partial^{e_j} p_{0,1}(\zeta_{s,j}(t),s_1)dt.
\]
Therefore
\begin{align*}
    \nu(s)
    &=\int p_{0,1}(\zeta_{s,j}(t_0),s_1)\phi(s_1){\rm d} s_{1} + \int \int_{t_0}^{s(j)} \partial^{e_j} p_{0,1}(\zeta_{s,j}(t),s_1)\phi(s_1) dt{\rm d} s_{1}\\
    &=\int p_{0,1}(\zeta_{s,j}(t_0),s_1)\phi(s_1){\rm d} s_{1} +\int_{t_0}^{s(j)}   \int \partial^{e_j}p_{0,1}(\zeta_{s,j}(t),s_1)\phi(s_1) {\rm d} s_{1} dt
\end{align*}
where the changing the order of integration is validate by Fubini's theorem under condition (d). This condition also indicates the inner integral is a continuous function of $t$, and hence $D^{e_j}\nu(s)$ exists and
\begin{align*}
   D^{e_j}\nu(s)
    &=\int \partial^{e_j}p_{0,1}(\zeta_{s,j}(s(j)),s_1)\phi(s_1) {\rm d} s_{1}=\int \partial^{e_j}p_{0,1}(s,s_1)\phi(s_1) {\rm d} s_{1}.
\end{align*}
Because $\partial^{e_j}p_{0,1}(s,s_1)$ is continuous and has integrable derivatives with respect to $s$ due to condition (d), the partial derivative $\partial^{e_j}\nu(s)$ is also continuous for all $j=1,\cdots,d$. This indicates $\nu(s)$ is differentiable. Similarly, we can establish the higher order differentiability of $\nu(s)$ through induction on the order.
\Halmos
\endproof
\end{APPENDICES}

\bibliographystyle{informs2014}  
\bibliography{main}

\end{document}